# Exploring Pedagogical Content Knowledge of Physics Instructors and Teaching Assistants Using the Force Concept Inventory

Alexandru Maries and Chandralekha Singh

*Department of Physics and Astronomy, University of Pittsburgh, Pittsburgh, PA 15260*

**Abstract**

The Force Concept Inventory (FCI) has been widely used to assess student understanding of introductory mechanics concepts by a variety of educators and physics education researchers. One reason for this extensive use is that many of the items on the FCI have strong distractor choices that correspond to students' alternate conceptions in mechanics. Instruction is unlikely to be effective if instructors do not know the common alternate conceptions of introductory physics students and explicitly take into account students' initial knowledge state in their instructional design. Here, we discuss research involving the FCI to evaluate the pedagogical content knowledge of both instructors and teaching assistants (TAs) of varying teaching experience. For each item on the FCI, the instructors and TAs were asked to identify the most common incorrect answer choice of introductory physics students. We also discussed the responses individually with a few instructors. Then, we used the FCI pre-test and post-test data from a large population (~900) of introductory physics students to assess the pedagogical content knowledge of physics instructors and TAs. We find that while both physics instructors and TAs, on average, performed better than random guessing at identifying introductory students' difficulties with FCI content, they did not identify many common difficulties that introductory physics students have, even after traditional instruction. Moreover, the ability to correctly identify students' difficulties was not correlated with the teaching experience of the physics instructors.

## Introduction: Background on past research involving the Force Concept Inventory

The Force Concept Inventory (FCI) is a multiple choice survey developed in 1992 by Hestenes, Wells and Swackhammer [1] and later revised [2] after many early observations made by Halloun and Hestenes [3] and other physics education researchers [4-7] that many students enter and leave physics classes with conceptions that are not consistent with the scientifically accepted concepts taught in the physics classes. The FCI was designed to assess student understanding of the fundamental mechanics concepts related to force and motion and has been widely used for this purpose by many educators and physics education researchers. Similar assessments in mechanics have been designed for the same purpose by other physics education researchers [8-12]. Although the conclusion that the FCI consistently measures Newtonian thinking was subject to some debate [13-16], the general consensus is that the FCI score is a good indicator of Newtonian thinking [17-20]. Some researchers have investigated the validity of the items on the FCI using Item Response Analysis [17-19]. Morris et al. [18] have argued that Item Response Analysis can be used to identify answer choices which do not discriminate between students in different ability groups (ability was mainly defined by them using FCI scores). Their analysis was also used to investigate student performance in more detail and gain some insights into student difficulties for some of the items on the FCI. Other researchers studied the FCI using the Rasch model [20] and concluded that the FCI "has succeeded in defining a sufficiently uni-dimensional construct for each population" (non-Newtonian and predominantly Newtonian). The analysis by Planinic et al. suggested that "the items in the test all work together and there are no grossly misfitting items which would degrade measurement" [20].

The FCI has played a key role in convincing many educators that traditional teaching methods which are primarily lecture oriented and do not engage students in the learning process actively do not promote conceptual and functional understanding [21-23]. Several studies have demonstrated that many students enter and leave introductory physics courses with the same alternate conceptions that are inconsistent with the accepted scientific ways of reasoning. Indeed, the use of the FCI in traditionally taught classes (even those taught by popular instructors) gave an impetus to the field of physics education research (PER) as educators increasingly realized that traditional methods were not working as intended, and consequently they began to develop and evaluate instructional strategies designed to promote functional understanding of physical phenomena [21-24]. The FCI has often been used to assess whether a particular instructional strategy is effective in promoting conceptual and functional understanding. Hake [21] used the FCI for this purpose and found that courses that make use of research based instructional approaches such as collaborative peer instruction [25-27], modeling [28-30], concept tests [24], microcomputer-based labs [31-33], active-learning problem sheets (ALPS) [34,35] and others [36,37] result in higher normalized gains on the FCI than courses which employ traditional methods such as standard lectures. The average normalized gain is defined as the ratio of the change in the average post-test score (after instruction of Newtonian concepts) with respect to the average pre-test score (before instruction of Newtonian concepts) to the average maximum possible change from the average pre-test score, i.e., average gain $<g>$ = ($<$post percent$>$−$<$pre percent$>$)/(100%−$<$pre percent$>$). Hake's study was performed on more than six thousand students and included both college and high-school classes.

The FCI has also been used to explore gender differences in understanding of Newtonian concepts related to force and motion [38-41]. Typically, in a particular introductory course, males outperform females on the FCI. However, in other course assessment measures such as the final exam, the males and females exhibit comparable performance. The gender gap observed on the FCI can be effectively reduced [38,40], although not necessarily removed [41], through PER based teaching strategies including but not limited to peer instruction, cooperative problem solving or using tutorials such as *Tutorials in Introductory Physics* by the University of Washington group. Other researchers have argued that the worse performance of females on the FCI can be partly attributed to the context of the questions which is mostly masculine and/or abstract [42]. Previous research indicated that females are more successful when questions are phrased using real-life contexts [43], therefore McCullough developed a "gender" version of the FCI [44] in which the items were rephrased from formal or male-oriented contexts to daily-life and female-oriented contexts. McCullough showed [42,44] that there was significant context dependence in the performance of both males and females for some questions. However, on individual questions different trends were observed  (e.g., male performance improved and female performance declined on some revised questions, performance of both stayed the same on some, and females improved and males declined on other revised questions, etc.) Overall, there was a decline in the overall performance of males, but the performance of females stayed about the same. Other researchers [45] have used differential item functioning to investigate whether some questions favor one gender over another (i.e., if a male student is statistically more likely than a female student of the same ability to answer a question correctly) and concluded that five questions may have a gender bias. Context dependent performance on FCI questions was also investigated by Dancy [46] who developed an animated version of the FCI and found differences in performance on seven questions.

Other researchers have used the FCI to investigate correlations between FCI scores and various other indicators of student performance: normalized gain on the FCI [47], problem solving ability [48], scientific reasoning ability [47,49], mathematics preparation [50], SAT scores [51], representational consistency [52], etc. In almost all these instances, significant positive correlations were found.

The FCI has often been administered by physics education researchers and curriculum developers as a pre-test to determine what initial knowledge students bring to the learning of physics. Knowing the initial knowledge state of students is important because instructional tools and pedagogies can be designed to take advantage of the knowledge resources students have and to effectively address the alternate conceptions which are not consistent with the accepted scientific way of reasoning about physical phenomena. In addition, the FCI has been administered as a post-test, e.g., to determine what concepts are difficult for students even after instruction and how effective instruction was at addressing student difficulties.

## Focus of our Research: Pedagogical Content Knowledge related to FCI

This research explores the pedagogical content knowledge of instructors of varying teaching experience and graduate teaching assistants (TAs) enrolled in a TA training course using the FCI. For each item on the FCI, the instructors and TAs at the University of Pittsburgh (Pitt) were asked to identify the most common incorrect answer choice of introductory physics students. We also discussed the responses individually with a few instructors and had a discussion with the TAs.

Pedagogical content knowledge (PCK) was defined by Shulman [53,54] as the subject matter knowledge *for teaching* and many researchers in K-16 education have adapted this construct. According to Shulman, PCK is a form of practical knowledge used by experts to guide their pedagogical practices in highly contextualized settings. In addition to the knowledge of the most useful forms of representation of the topics, use of powerful analogies, illustrations and examples, etc., Shulman included in pedagogical content knowledge, "understanding of the conceptions and preconceptions that students bring with them to the learning of those most frequently taught topics and lessons" [54]. Our research presented here explores the PCK of the physics instructors and graduate TAs with regard to their knowledge of introductory physics students' alternate conceptions related to force and motion. In particular, we investigate whether instructors and graduate students are able to identify the common alternate conceptions of students on individual items in the FCI. Knowledge of the alternate conceptions which are inconsistent with the scientifically accepted way of reasoning about the concepts can be helpful in devising the curricula and pedagogical strategies to improve student understanding. Much physics education research has been devoted to devising and assessing such strategies.

We note that, in order to carry out the research related to a particular aspect of PCK dealing with students' alternate conceptions, we needed introductory student data for each answer choice on individual items on the latest version of the FCI from large populations of students. It is most appropriate to analyze instructor and graduate student PCK data at Pitt by comparing it to introductory physics students' FCI data at the same institution, which is a large, typical state related university of about 18,000 undergraduate students. Therefore, data were collected over a few years both in pre-tests (before instruction) and post-tests (after instruction) from about 900 algebra-based students and over 300 calculus-based students. The courses were all taught using traditional instructional methods at Pitt. These data were used to determine the common student alternate conceptions related to each item on the FCI and thus to assess the PCK of physics instructors and graduate students enrolled in a TA training course related to their knowledge of common student difficulties.

**Primary research questions – pedagogical content knowledge related to FCI**

In order to investigate the PCK of physics instructors and first year physics graduate TAs enrolled in a TA training course related to FCI content, specific research questions were developed as follows:

**I. Here we summarize the research questions related to the performance of instructors and graduate student TAs at identifying introductory physics students' alternate conceptions on the FCI and later expand upon them in the methodology section.**

I. a. Does teaching experience influence the ability to identify introductory students' alternate conceptions?

I. b. Are American physics graduate students, who have been exposed to undergraduate teaching in the United States, better at identifying introductory students' alternate conceptions than foreign physics graduate students?

I. c. To what extent do instructors and/or graduate students identify 'strong' student alternate conceptions compared to 'medium' level ones?

I. d. Do graduate students identify student alternate conceptions more often when working in groups of two or three than when working individually (i.e., do discussions improve graduate students understanding of introductory students' alternate conceptions?)?

I. e. To what extent do instructors/graduate students identify specific alternate conceptions of introductory physics students? Is their ability to identify these alternate conceptions context dependent?

**II. Secondary research questions – introductory student FCI performance**

In order to answer the primary research questions, we needed data on the performance of students on individual items on the FCI (the revised version of the test from 1995). Therefore, we collected FCI data from about 900 students in algebra-based and more than 300 students in calculus-based introductory physics courses at Pitt. Subsequently, the following secondary research questions emerged related to analysis of introductory student performance on individual questions on the FCI and comparison of the pre-test and post-test data for both algebra-based students and calculus-based students.

**II. a. Which questions on the FCI pose significant challenges for students?**

**II. b. Are there any questions on the FCI in which there is little improvement (small normalized gain) from pre-test to post-test?**

**II. c. Are there any shifts in the most common alternate conceptions from the pre-test to the post-test?**

**II. d. On which questions do calculus-based students perform better than algebra-based students by 20% or more? Are there any questions in which the most common alternate conceptions of algebra-based students are different from the most common alternate conceptions of calculus-based students?**

## Methodology

Thirty physics instructors, and 25 first year physics graduate students, enrolled in a semester-long TA training course at Pitt participated in this study. They were given the FCI survey and for each question, they were asked to identify which one of the four *incorrect* answer choices, in their view, would be most commonly selected by introductory physics students after instruction in relevant concepts if the students did not know the correct answers. We note that many FCI questions have incorrect answer choices which include explicit reasons for selecting them. Therefore, we did not ask for explanations for why the instructors and graduate TAs who participated in this study felt that a certain incorrect answer would be the most commonly chosen incorrect answer by an introductory student who did not know the correct answer. We did discuss the reason for why they thought that introductory physics students will select a particular incorrect answer choice with some faculty members in a one on one discussion and with graduate TAs in the TA training class for the questions in which the incorrect answers do not have an explanation in the FCI. We note that the task given to the instructors and graduate students was framed such that they had to identify the most common incorrect option for each multiple choice question that introductory physics students would select after instruction if they did not know the correct answer (rather than the most common alternate conceptions of introductory students before instruction), because individual discussions with some faculty members (who had taught introductory physics) before giving them the task indicated that they felt that they had no way of knowing the "pre-conceptions" of introductory physics students. Their reluctance to contemplate introductory physics students' preconceptions about force and motion before instruction motivated us to ask them to identify the most common incorrect answer choice for each question if the student did not know the correct answer after instruction in relevant concepts. Although asking them to identify the most common alternate conception in a post-test made it easier for them to complete the task, some faculty members who participated in the study were concerned about their ability to identify students' difficulties and explicitly noted that they have no way of knowing the most common difficulty of introductory students for each question.

We also note that it does not make a significant difference whether the question is phrased to the instructors and graduate students about introductory physics students' difficulties with each question in the post-test or pre-test because the common alternate conceptions of introductory students rarely changed after traditional instruction. Instead, typically, fewer students held the common alternate conceptions (this was found to be true when we compared the pre-test and post-test data of introductory students). Therefore, the performance of experts (instructors and graduate TAs) at identifying these alternate conceptions provides an indication of their knowledge of the initial state of introductory students.

The instructors were asked to complete the task at their convenience. Also, the task was originally given to 33 physics instructors at Pitt but three of them did not complete the task in a reasonable amount of time even after multiple reminders. The instructors varied widely in terms of introductory physics teaching experience. In particular, some instructors were relatively new and had only taught introductory courses a few times, while others were emeritus professors who had not taught for a while (but had taught a long time ago) and yet others were instructors who teach introductory physics courses on a regular basis. After the instructors had completed the task, we discussed the reasoning for their responses individually with some of them, especially for the questions in which the reasoning was not explicitly provided.

As noted earlier, the FCI was also given to twenty-first year physics graduate students who were enrolled in a semester long TA training course toward the end of the TA training course. A majority of the graduate students were at the time teaching introductory physics recitations or introductory labs for the first time except a few graduate students who were on fellowships and had not yet taught introductory physics. The graduate students in the TA training class were first asked to solve the FCI (by determining the correct answers), then, similar to the instructors, they were asked to identify the most common incorrect answer choice of introductory physics students who did not know the correct answer for each question in a post-test. Lastly, the graduate students repeated this latter task of identifying the most common introductory student difficulties for each question while working in groups of two or three. The graduate student population at Pitt is consistent with that of a typical research focused state university and the nationality of the graduate students varied: nine graduate students were from the United States, nine were from China and the other seven were from other countries (Asian and European). After the graduate students completed the task, there was a discussion in the TA training course about the FCI related PCK task and why knowing the difficulties of the students is critical for teaching and learning to be effective in general.

In order to determine how well instructors and first year physics graduate students at the end of a TA training course identified introductory physics students' common alternate conceptions related to the FCI, we used post-test data from several algebra-based introductory physics courses (about 900 students, included in the appendix) collected over several years. These are the data we mainly used to carry out research on FCI related PCK of physics instructors and graduate students. We note that we collected the FCI pre-test and post-test data for both algebra-based and calculus-based courses and we will briefly compare the algebra-based and calculus-based classes in the results section. The introductory physics FCI pre-test and post-test data were collected at Pitt, which is a large research focused state university. All classes from which these data was collected were taught in a traditional manner and the average unmatched (all students who took the pre-test and post-test were included regardless of whether they took both the pre-test and post-test) normalized gain was 0.26 (not much different from the matched normalized gain), a typical gain for courses that do not employ PER based instructional strategies as reported by Hake [21].

In order to compare the FCI related PCK performance of the physics instructors with that of the graduate students (and also to compare the FCI related PCK performance of different subgroups of instructors/graduate students), scores were assigned to each instructor/graduate student. An instructor/graduate student who selected a particular incorrect answer choice as the most common incorrect choice in a particular question received a PCK score which was equal to the fraction of introductory students who selected that particular incorrect answer choice. If an instructor/graduate student selected the correct answer choice as the most common incorrect answer (which happened for a few questions), they were assigned a score of zero because they were explicitly asked to indicate the *incorrect* answer choice which is most commonly selected by introductory students if they did not know the correct answer. For example, in question 2, the fractions of algebra-based students who selected A, B, C, D and E are 0.44, 0.25, 0.06, 0.21 and 0.04, respectively (see Table A1). Answer choice A is correct, thus, the score assigned to instructors or graduate students for each answer choice if they selected it as the most common incorrect answer would be 0, 0.25, 0.06, 0.21 and 0.04 (A, B, C, D and E). The total score an instructor/graduate student would obtain on the task for the entire FCI can be obtained by summing over all of the questions. In order to determine whether the instructors/graduate students performed better than random guessing on the FCI related PCK task, a population of random guessers was generated. The population was generated by choosing N=24 'random guessers' in order to have a reasonable group size when performing *t*-tests [55]. Random guessing on this task would correspond to choosing one of the four incorrect answer choices for each question with equal probability (25%). Therefore, one quarter of the random guessers always selected the first incorrect answer choice, one quarter selected the second incorrect answer choice, etc. Since the instructors and graduate students were not told the correct answers before they performed the FCI related PCK task, random guessing would not perfectly correspond to selecting one of the four incorrect answer choices with equal probability. For a particular question, there is a small probability that an instructor/graduate student does not know the correct answer. However, our data indicate that this probability is very small (see table A1 included in the appendix). Moreover, since for a given question, one quarter of the random guessers selected each of the four incorrect answer choices, one can calculate a mean and a standard deviation that can be used to perform comparison with the graduate student scores. Furthermore, our choice of random guessers maximizes the standard deviation.

We note that our approach used to determine the PCK score related to FCI appropriately weighs the responses of instructors/graduate students by the fraction of introductory students who selected a particular incorrect response although the instructors and grad students were asked to select the most common incorrect answer choice for each question. The total PCK score can be calculated mathematically for the instructors (and similarly for graduate students and 'random guessers') if we define indices $i$, $j$ and $k$ that correspond to the following:

- $i$: index of instructor (30 instructors; it takes values from 1 to 30);
- $j$: FCI question number (30 questions; it takes values from 1 to 30);
- $k$: incorrect answer choice number for each question (4 incorrect answer choices; it takes values from 1 to 4).

Then, we let $F_{jk}$ be the fraction of introductory physics students who selected incorrect answer choice $k$ on item $j$ (e.g. $F_{21} = 0.44$, $F_{22} = 0.06$, $F_{23} = 0.21$, $F_{24} = 0.04$). We let $I_{ijk}$ correspond to whether instructor $i$ chose incorrect answer choice $k$ on item $j$ (for a given $i$ and $j$, $I_{ijk}$=1 only for the incorrect answer choice $k$, selected by instructor $i$ on item $j$, otherwise $I_{ijk}$=0). Then, the PCK score of the $i$th instructor on item $j$ (referred to $I_{ij}$) is: $I_{ij} = \sum_{k=1}^{4}(I_{ijk} \cdot F_{jk})$. Then, the total PCK score of the $i$th instructor ($I_i$) on the whole survey can be obtained by summing over all of the questions:

$$I_i = \sum_{j=1}^{30} I_{ij} = \sum_{j=1}^{30}\left[\sum_{k=1}^{4}(I_{ijk} * F_{jk})\right].$$

Also, the PCK score of all of the instructors on item $j$ (referred to as $\overline{I_j}$) can be obtained by summing over the instructors:

$$\overline{I_j} = \sum_{i=1}^{30} I_{ij} = \sum_{i=1}^{30}\left[\sum_{k=1}^{4}(I_{ijk} * F_{jk})\right].$$

A similar approach can also be adopted for the graduate students ($GS_{ij}$ – PCK score of the $i$th graduate student on item $j$; $GS_i$ – PCK score of the $i$th graduate student on the whole survey; $\overline{GS_j}$ – PCK score of all graduate students on item $j$) and for random guessers ($RG_{ij}$ – PCK score of $i$th random guesser on item $j$; $RG_i$ – PCK score of $i$th random guesser; $\overline{RG_j}$ – PCK score of random guessers on item $j$). The PCK scores of each ($i$) instructor/graduate student/random guesser ($I_i$, $GS_i$, $RG_i$ as described above) were used to obtain averages and standard deviations in order to perform $t$-tests to compare the FCI related PCK performance of instructors with that of the graduate students and random guessers on the whole survey (and to compare different subgroups of instructors and graduate students). In order to compare the PCK performance of these different groups on individual items, the averages and standard deviations of the PCK scores on that particular question (e.g., for question $j$ on the FCI: $I_{ij}$, $GS_{ij}$, $RG_{ij}$) were used to perform $t$-tests.

**Methodology for answering the research questions**

**Primary research questions – pedagogical content knowledge related to the FCI**

**I. Performance of instructors and graduate students at identifying introductory physics students' alternate conceptions as they are exhibited by the FCI**

The researchers analyzed whether instructors and/or graduate students performed better at identifying introductory students' alternate conceptions than random guessers by performing statistical analysis.

**I. a. Does teaching experience influence the ability to identify introductory students' alternate conceptions?**

In order to answer this question, we compared the average PCK score of all instructors with the average PCK score of all graduate students and also compared the PCK scores of instructors who had recently taught introductory mechanics (either algebra-based or calculus-based) with those who had not taught introductory mechanics recently.

The PCK scores of instructors (all of whom had taught some introductory physics course in the near or distant past and several had taught it many times) were compared with the PCK scores of graduate students enrolled in the TA training course (at the end of the course) who had never taught an introductory physics course as lecturers before. All of the graduate students were at the time in their first semester in physics graduate school and most were doing a teaching assistantship for the first time. Since the teaching experience as lecturer of the graduate students was very limited compared to the teaching experience of most of the instructors who had taught some introductory courses, this comparison may provide some indication for whether teaching experience as lecturer influences the ability to identify student alternate conceptions. We note however, that the first year physics graduate students in the TA training course had taken introductory physics only a few years prior to the study as undergraduates and a majority of them were TAs for introductory recitations and laboratories, graded homework, quizzes and exams and held office hours in the resource room at Pitt to help introductory students throughout the semester. These experiences may help the graduate students identify introductory students' alternate conceptions and therefore, it is difficult a priori to predict how they will perform compared to the instructors (most of whom did minimal grading and had minimal direct contact with students in the large introductory classes) regardless of the fact that instructors had significantly more independent classroom teaching experience.

Therefore, we also compared the FCI related PCK scores of instructors who had taught introductory mechanics recently with those who had not taught it or had not taught it in the last seven years. Half of the instructors who participated in this study had taught introductory algebra-based or calculus-based mechanics courses at least a few times in the past seven years, while the other half had not taught these courses or taught them more than seven years prior to the study. This analysis was designed to investigate if recent teaching experience in introductory algebra-based or calculus-based mechanics courses played a role in the instructors' ability to identify introductory students' alternate conceptions about force and motion.

### I. b. Are American graduate students, who have been exposed to teaching in the United States, better at identifying student alternate conceptions than foreign graduate students?

Out of the twenty-five first year graduate students who participated in this study, nine were American, nine were Chinese and seven were from other foreign countries (Asia and Europe). The PCK scores of three groups of graduate students were compared (American, Chinese and other foreign students). The reason we divided the graduate students in three groups is because the American graduate students were exposed to teaching in the United States as opposed to the foreign students, who were not exposed to US teaching practices before graduate school and many were taught physics in their own native languages. The nine Chinese graduate students were placed in a separate group because, although they fit the category of foreign graduate students, it is possible that their backgrounds are different from the backgrounds of most of the other foreign graduate students.

### I. c. Are instructors and/or graduate students better at identifying 'strong' student alternate conceptions than 'medium' level ones?

Many of the questions on the FCI contain strong distractor choices that are selected by a large number of introductory students even in a post-test. The researchers determined that an incorrect answer choice can be attributed to a 'strong' student alternate conception if more than 1/3 of introductory algebra-based students selected that answer choice. An incorrect answer choice was considered to be 'medium' level distractor choice if between 19% and 34% of the students selected that answer choice (initially, the lower cutoff was chosen to be 20%, but there were three questions on the FCI in which 19% of introductory students selected an incorrect answer choice, two of which were deemed interesting by the researchers, thus 19% was selected to be the lower cutoff).

In order to answer whether physics instructors or graduate students are better at identifying strong alternate conceptions than medium level ones, we compared how often instructors or graduate students performed better than random guessing on questions which contained strong alternate conceptions with how often they performed better than random guessing on questions which contained medium level alternate conceptions.

### I. d. Do graduate students identify student alternate conceptions more often when working in groups of two or three than when working individually? (i.e., do discussions improve graduate students understanding of introductory students' alternate conceptions?)

Previous studies have found that student discussions improve performance on conceptual examinations [24, 56]. Mazur's Peer Instruction [24] approach has been developed because student discussions tend to converge to the correct answers rather than the incorrect answers. In particular, research suggests that if two students individually select different answers and one of them is correct, the student with the correct answer is more likely to convince the student with the incorrect answer through a discussion than otherwise. We investigated whether the same is true when the discussions are among graduate students and are centered on introductory students' alternate conceptions.

The graduate students completed three tasks related to the FCI in a two hour long TA training class toward the end of the semester: first they were asked to provide the correct answers to the FCI, second, they were asked to identify the most common incorrect answer choice of introductory physics students who did not know the correct answer for each question in a post-test and third, they repeated this PCK task in groups of two or three. It was investigated whether discussions among graduate students improved their knowledge of introductory student alternate conceptions. Two factors would indicate that discussions improve graduate students' understanding of introductory students' alternate conceptions: 1) better FCI related PCK performance and 2) convergence to a more common introductory student alternate conception. The second factor warrants further explanation: if in the individual PCK task, two graduate students selected two different incorrect answer choices (that they thought would be most common among introductory students who did not know the correct answer), and at least one of the incorrect answer choices is connected to a common student alternate conception, we investigated how often the two graduate students agreed on the incorrect answer choice which is selected by more introductory students. In order to answer this question, we identified all the instances in which two (or three) graduate students who selected different incorrect choices in the individual PCK task, while working in a group, agreed on one of the incorrect answers. Then, we determined how often the incorrect answer selected in the group PCK task was more common (by 5% or more) among introductory students than the other answers selected by the graduate students in the individual PCK task.

**I. e. To what extent do instructors/graduate students identify particular introductory student alternate conceptions? Is their ability to identify these alternate conceptions context dependent?**

These questions were answered by identifying particular alternate conceptions (e.g., constant force implies constant velocity) in different questions and analyzing instructor/graduate student PCK performance in identifying these common alternate conceptions in different questions.

## II. Secondary research questions – introductory student FCI performance

**II. a. Which questions on the FCI pose significant challenges for students?**

This question was answered while analyzing the PCK performance of instructors and graduate students at identifying students' alternate conceptions because this analysis was restricted to the alternate conceptions held by at least 19% of introductory students. For each alternate conception, the question in which it appears and the percentage of introductory students who hold the particular alternate conception was identified.

**II. b. Are there any questions on the FCI in which there is very little improvement from pre- to post-test?**

Introductory student performance in a post-test is not the sole indicator of how difficult a question is. If the percentage of introductory students who answer a question correctly does not change significantly after instruction in the relevant concepts, it is an indicator of the difficulty of the question regardless of the percentage correct in the post-test. The determination of questions with "little" improvement from pre-test to post-test was done based on two criteria: the average normalized gain (see Table A3) and improvement in the percentage of students holding an alternate conceptions. For normalized gain, the questions were ordered from lowest to highest and the researchers determined that "little" improvement occurred in the bottom 1/3 of the questions. For the second criteria, it was considered that "little" improvement occurred in questions in which the improvement in the percentage of students who hold the most common alternate conception is less than 5%.

**II. c. Are there any shifts in alternate conceptions from the pre-test to the post-test?**

Previous research has found that many students enter introductory mechanics classes with naïve interpretations of real world phenomena that are inconsistent with physics principles [1-10]. One may expect that after instruction, the performance of introductory students on individual items would improve and the incorrect answers which were selected most commonly in the post-test would largely remain the same as the ones that would be selected most commonly in the pre-test, except by smaller percentages of students in the post-test. However, students might shift from one incorrect answer choice in the pre-test to another incorrect answer choice in the post-test. For example, in question 5 (identify all the forces that act on a ball while it is moving in a frictionless, circular channel), before instruction, many students do not know that the channel exerts a force on the ball, but know about the force of gravity and think that in order for an object to be moving in a certain direction, a distinct force must be acting on it in the direction of motion. It is possible that after instruction, most of these students learn that the channel must exert a force on the ball, but do not abandon the idea that a distinct force must exist in the direction of motion. The post-test response based upon these notions would still be incorrect; however, the alternate conception will now be different than in the pre-test.

In order to determine whether algebra-based introductory students hold different alternate conceptions after instruction compared to before instruction we analyzed questions which had two or more common alternate conceptions either in the pre-test or the post-test. In these questions, it was considered that a shift occurred if the following changes transpired from the pre-test to the post-test: 1) the percentage of introductory students who selected one of these incorrect answer choices decreased (by 10% or more) while the percentage of introductory students who selected the other incorrect answer choice(s) remained the same or increased or 2) the percentage of introductory students who selected one of the incorrect answer choices remained the same, while the percentage of introductory students who selected the other incorrect answer choice(s) increased.

### II. d. On which questions do calculus-based students perform better than algebra-based students? Are there any questions in which the alternate conceptions of algebra-based students are different from the alternate conceptions of calculus-based students?

Previous research has found that students in calculus-based classes perform better than students in algebra-based classes on the FCI [21,57] and other conceptual assessments [8,58]. However, it is possible that on some FCI questions the differences are less pronounced than in others. We investigated on which questions on the FCI the calculus based students performed better than the algebra based students and on which questions the differences were small. In addition, we investigated whether there were any questions for which the most common alternate conceptions of algebra-based students were different from the common alternate conceptions of calculus-based students.

### Results

Many instructors and graduate students noted that the task of thinking from a student's point of view was challenging; some even confessed that they did not feel confident about their performance in identifying the most common incorrect answers. Also, the task was posed as the identification of the most common incorrect answer of introductory physics students for each FCI question *after* instruction if students did not know the correct answer. Thus, the primary data analysis in this section involves comparison of the instructors' and graduate students' responses with introductory physics responses on each FCI question *after* instruction. However, our analysis revealed that the introductory students' alternate conceptions are generally the same, except more pronounced before instruction compared to after instruction.

**Primary research questions – pedagogical content knowledge related to the FCI**

**I. Performance of instructors and graduate students at identifying introductory physics students' alternate conceptions related to the FCI**

There are 24 questions on the FCI which reveal strong and/or medium alternate conceptions: items 2, 4, 5, 9 and 11-30. Analysis of the FCI related PCK of both instructors and graduate students was conducted on each of these questions and the results are displayed in Tables A1 and A2 (included in the appendix).

Table A1 shows the percentages of introductory physics students who selected different answer choices when asked to select the correct choice for each question, and instructors and graduate students who selected each answer choice for what would be the most common incorrect choice of introductory physics students if they did not know the correct answer on each of the 24 questions in which strong or medium level alternate conceptions were identified. Correct answers are indicated by the green shading in Table A1, strong student alternate conceptions (incorrect answer choices selected by more than 1/3 of the introductory students) are indicated by red shading and medium alternate conceptions are written in red. In addition, the second column (titled >RG) in Table A1 indicates whether instructors and/or graduate students performed better than random guessing. For each question, "I" in the second column of Table A1 indicates that instructors performed better than random guessing, "GS" indicates that graduate students performed better than random guessing and "I, GS" indicates that both instructors and graduate students performed better than random guessing in identifying introductory physics students' most common incorrect answer for a particular question. If the field in the second column of Table A1 is blank then neither instructors, nor graduate students performed better than random guessing.

Table A2 shows, for each question, the normalized average FCI related PCK scores of the instructors and graduate students. Their scores were normalized on a scale from zero to 100 because for each question on the FCI there is a minimum and a maximum possible score, which correspond to the smallest and largest fractions of introductory students who selected a particular incorrect answer choice. The normalization was done in the following manner: normalized score = 100 * (average PCK score – minimum possible score) / (maximum possible score – minimum possible score). The normalized PCK score is then zero if the instructors/graduate students obtained the minimum possible score and 100 if they obtained the maximum possible score. This also provides a way to compare the PCK performance in different questions which have different minimum and maximum possible PCK scores. Table A2 also shows the difficulty of each of these questions via the percentage of introductory algebra-based students who answered each question correctly in a post-test, normalized gain and strength of the alternate conception(s), i.e., medium level or strong. The questions which contained a strong alternate conception are indicated by the red shading and those which contained a medium level alternate conception are written in red. Also, the performance of instructors and graduate students is considered 'good' (and shaded green) if their normalized FCI related PCK score is more than 2/3 of the maximum possible score, 'medium' level (and shaded yellow) if their normalized score is between 1/2 and 2/3 of the maximum possible score and 'poor' (horizontal stripes) if their performance is less than 1/2 of the maximum possible score. Examination of Table A2 indicates that the strength of an alternate conception is not correlated with the PCK performance of instructors and/or graduate students. Table A2 shows that there are questions with strong alternate conceptions in which both instructors' and graduate students' PCK performance is poor, other questions with strong alternate conceptions in which their PCK performance is medium level and yet others in which their performance is good. A similar observation can be made for questions in which there is a medium alternate conception. These results are discussed in more detail in section I. c.

**I. a. Does teaching experience influence the ability to identify introductory students' alternate conceptions?**

In order to answer this question, one analysis involved comparison of the overall scores ($I_i$) of instructors, who on the average had significant experience teaching introductory physics courses as lecturers, with the scores of graduate students enrolled in the TA training course ($GS_i$), who had limited or no experience teaching introductory physics courses as lecturers. The maximum possible FCI related PCK score of instructors or graduate students on each question would be equal to the maximum fraction of introductory students who selected an incorrect answer choice. The maximum possible FCI related PCK score on the whole survey is 9.21 which is the sum of these fractions for all the questions. Table 1 shows that the average of instructors (68% of the maximum possible FCI related PCK score) and the average of graduate students (65% of the maximum possible FCI related PCK score) are very close. Also, $t$-tests revealed no significant difference between instructors and graduate students in terms of their FCI related PCK scores. Although, their overall PCK performance is the same, there were many differences observed in the performance of identifying specific student alternate conceptions. However, both the instructors and graduate students performed significantly better on the FCI related PCK task than random guessing (both p values when comparing instructors to random guessing and graduate students to random guessing were less than 0.001). We note that since the graduate students had taken introductory physics only four years prior to this study as an undergraduate and the vast majority of them were TAs in an introductory recitation or laboratory class, did weekly grading of quizzes, homework and exams and held office hours in which they helped introductory students individually, they may identify with introductory physics students' difficulties related to FCI concepts.

**TABLE 1.** Numbers of instructors/graduate students/random guessers, averages and standard deviations (Std. dev.) for the FCI related PCK scores obtained (in determining student alternate conceptions on the FCI) out of a maximum of 9.21.

|  | N | Average | Std. dev. |
|---|---|---|---|
| Instructors | 30 | 6.25 | 0.90 |
| Graduate students | 25 | 6.01 | 0.78 |
| Random guessing | 24 | 3.71 | 0.93 |

We also investigated whether recent teaching experience in algebra-based or calculus-based introductory mechanics course was related to the ability of instructors to identify students' alternate conceptions that emerge in the FCI. The average of the instructors who had taught introductory mechanics courses in the past seven years was nearly identical to the average of instructors who had not taught those introductory physics courses recently (see Table 2). It appears that recent teaching experience of these instructors in introductory mechanics is not related to their ability to identify introductory students' alternate conceptions.

**TABLE 2.** Numbers of instructors who had taught and who had not taught introductory mechanics in the past seven years, their averages and standard deviations (Std. dev.) for the scores obtained for determining students' alternate conceptions on the FCI out of a maximum of 9.21.

|  | N | Average | Std. dev. |
|---|---|---|---|
| Have taught in the past 7 years | 15 | 6.33 | 0.77 |
| Have not taught in the past 7 years | 15 | 6.17 | 1.03 |

## I. b. Are American graduate students, who had been exposed to teaching in the United States, better at identifying student alternate conceptions than foreign graduate students?

Our analysis suggests that it was not the case that American graduate students performed better than the others. In particular, the averages of these three groups of graduate students (American, Chinese, other foreign) were very similar as shown in Table 3. Statistical analyses using $t$-tests are not appropriate here because the group sizes are small, but it does appear that the averages are not very different. The Chinese students were placed in a separate group because they comprised more than half of the foreign graduate students and we did not want the performance of foreign graduate students to be skewed because of this.

**TABLE 3.** Numbers of American/Chinese/Other foreign graduate students, their averages and standard deviations (Std. dev.) for the scores obtained in determining student alternate conceptions on the FCI out of a maximum of 9.21.

|  | N | Average | Std. dev. |
|---|---|---|---|
| American | 9 | 6.20 | 0.70 |
| Chinese | 9 | 6.04 | 0.76 |
| Other foreign | 7 | 5.71 | 0.91 |

## I. c. Are instructors and/or graduate students better at identifying 'strong' student alternate conceptions than 'medium' level alternate conceptions?

There are 11 questions on the FCI in which at least 1/3 of the introductory physics students selected a particular incorrect answer choice (see Table A1). The instructors' PCK score was better than random guessing on eight of these questions (73%) while the graduate students' PCK score was better than random guessing on five (45%) of these questions. In the other 13 questions (which contained 'medium level' alternate conceptions, i.e. conceptions held by 19%-33% of introductory students in a post-test), both instructors' and graduate students' PCK scores were better than random guessing on seven of them (54%). These numbers are too small to perform meaningful statistics, but it appears that instructors identified 'strong' misconceptions somewhat better than 'medium level' ones (73% compared to 54%), whereas for graduate students, the difference is minor (45% as compared to 54%).

**I. d. Do graduate students identify student alternate conceptions more often when working in groups of two or three compared to when working individually? (i.e., do discussions improve graduate students understanding of student alternate conceptions?)**

    1) Graduate student FCI related PCK performance is better when they work in groups compared to when they work individually.

Table 4 shows the graduate students' FCI related PCK performance when they worked individually and in groups of two or three. A *t*-test shows that the group performance is better than the individual performance (p = 0.040).

**TABLE 4.** FCI related PCK performance of graduate students in the individual and in the group PCK tasks: number of graduate students/groups (N), averages (Avg.) and standard deviations (Std. dev.)

| Graduate students' FCI related PCK performance | | | |
|---|---|---|---|
| Individual | **N** | **Avg.** | **Std. dev** |
| | 25 | 6.01 | 0.78 |
| Group | **N** | **Avg.** | **Std. dev** |
| | 12 | 6.59 | 0.79 |

    2) Discussions among graduate students often tend to lead them to agree on a more common introductory student alternate conception

There were 98 instances in which two or three graduate students who did not all select the same incorrect answer choice in the individual PCK task, when working in groups, converged to one of their original answers pertaining to introductory students' common difficulties. In 73 of those instances (74%) the graduate students converged to the 'better' option (i.e., the more common incorrect answer choice of introductory students by 5% or more) and in 25 of those instances (26%), they did not converge to the 'better' answer choice. It therefore appears that discussions among graduate students tend to lead them to agree on a more common introductory student alternate conception.

**I. d. To what extent do instructors/graduate students identify particular introductory student alternate conceptions? Is their ability to identify these alternate conceptions context dependent?**

These questions were answered by identifying student alternate conceptions, the questions in which these alternate conceptions are connected to incorrect answer choices and analyzing the FCI related PCK performance of instructors and graduate students in those questions. Similar alternate conceptions were grouped whenever it was deemed appropriate by the researchers (e.g. alternate conceptions related to Newton's third law, alternate conceptions related to particular tasks, such as identifying all the distinct forces that act on an object, etc.) and, if a particular alternate conception appeared in more than one context, it was investigated whether instructors and/or graduate students performed better at identifying it in some contexts than in other contexts. For multiple choice questions, the context is comprised of both the physical situation presented in the problem and the answer choices, because different answer choices can change the difficulty of a question. For example, a multiple-choice question is easier for introductory students if the incorrect answer choices are not chosen to reflect common student difficulties, and are challenging for students when they are chosen to reflect common difficulties [2-3].

**I. d. i) Newton's third law: The alternate conceptions of students related to Newton's third law and the performance of both instructors and graduate students in identifying the most common alternate conceptions are both context dependent.**

**TABLE 5.** Introductory students' alternate conceptions related to Newton's 3$^{rd}$ law, questions in which these alternate conceptions arise (FCI item #), percentage of introductory students who answer the questions incorrectly in the pre-test (% overall incorrect pre) and in the post-test (% overall incorrect post), incorrect answer choices on each question which uncover these alternate conceptions (incorrect answer choices), percentage of introductory students who hold the alternate conceptions based on their selection of these answer choices in the pre-test (Intro stud. alt. pre) and in the post-test (Intro stud. alt. post.) and percentage of instructors (Ins.) and graduate students (GS) who identify them as the most common incorrect answer choices. For convenience, brief descriptions of the problems are given underneath.

| Introductory student alternate conceptions | FCI item # | % overall incorrect pre | % overall incorrect post | Incorrect answer choices | Intro stud. alt. pre. | Intro stud. alt. post | Ins. | GS |
|---|---|---|---|---|---|---|---|---|
| Newton's 3$^{rd}$: while both objects exert forces on one another, if both objects are active (i.e. collision), the larger object exerts the larger force; if only one is active (i.e. car pushing truck), the active object exerts a larger force on passive object than vice versa | 4 | 74% | 40% | A | 73% | 39% | 97% | 84% |
| | 15 | 75% | 56% | C | 61% | 48% | 60% | 40% |
| | 16 | 45% | 27% | C | 37% | 19% | 37% | 16% |
| | 28 | 76% | 41% | D | 61% | 32% | 38% | 52% |
| **Questions** | | | | | | | | |
| 4. Truck colliding with car. | | | | | | | | |
| 15. Car pushing truck and speeding up. | | | | | | | | |
| 16. Car pushing truck and moving at constant speed. | | | | | | | | |
| 28. Student "a" puts his feet on student "b" and pushes against student "b". | | | | | | | | |

Table 5 shows that in some contexts, introductory students hold alternate conceptions related to Newton's third law more strongly (questions 4, 15 and 28 for which at least 32% of introductory students hold an alternate conception) than in other contexts (question 16, in which only 19% of introductory students hold an alternate conception). Thus, these alternate conceptions are context dependent and they arise more often in certain contexts than in others. This is similar to the finding by Redish and Elby [59] that students can answer paired questions about the same concept differently in different contexts: in one context most students answer it correctly, while in another context the most common incorrect answer involves a common student alternate conception.

In addition, it appears that the FCI related PCK performance of both instructors and graduate students is also context dependent. For example, the vast majority of both instructors and graduate students identified the alternate conception related to Newton's third law in a typical context (question 4 – truck colliding with car – see Table 5), but they did not identify it as often in the other three contexts (question 15 – car pushing truck and speeding up, question 16 – car pushing truck at constant speed and question 28 – student "a" pushing student "b"). Also, in question 15, 10% of instructors and 12% of graduate students selected the correct answer choice as the most common incorrect answer choice selected by introductory students (see Table A1). It is likely that, in this context, they have the same alternate conception as introductory students, namely, that while the car is speeding up, it exerts a larger force on the truck than vice versa. In addition, as noted earlier, the graduate students were first asked to identify the correct answers on the FCI before performing the FCI related PCK task and 24% of them incorrectly selected this answer choice as the correct one (this was one of the two questions with the lowest graduate student performance when asked to select the correct answers for the FCI questions). The fact that even some experts hold this alternate conception after many years of practicing physics points out how strong this alternate conception is and how difficult it is to overcome it in this particular context. Question 16, although relatively easy for introductory students (73% of them answered it correctly in a post-test), revealed a medium level alternate conception, namely that the force the car exerts on the truck is larger than the force the truck exerts on the car. On the other hand, both instructors and graduate students performed very poorly on this question on the FCI related PCK task. In particular, a majority of them (60% of instructors and 76% of graduate students) selected answer choices B, D and E which were selected by only 8% of introductory students (see Table A1). Similarly, in question 28, many instructors and graduate students performed poorly on the PCK task of identifying the most common alternate conception and selected answer choice B (45% instructors and 36% graduate students) which stated that student "a" exerts a force on student "b", but student "b" does not exert a force on student "a". However, very few introductory students selected this answer choice (2%) and the vast majority of them knew that both students exert forces on one another (91% who selected answer choice C, D or E).

**I. d. ii) Identification of distinct forces: In the following questions which ask introductory students to identify all the distinct forces acting on an object, neither instructors nor graduate students identified the most common student alternate conceptions and many graduate students, and even more instructors selected answer choices which either ignored contact forces or all forces altogether, inconsistent with introductory students' most common incorrect answer choices.**

Table 6 shows that the majority of both instructors and graduate students are aware that introductory students have the alternate conception that moving objects are acted on by a distinct force in the direction of motion. However, in all these questions, many instructors and graduate students claimed that introductory students will not identify contact forces (normal and tension forces), and to a lesser extent they will not identify any forces (including the force of gravity) even in the post-test. However, contrary to what instructors and graduate students claimed, introductory students rarely selected answer choices which correspond to these alternate conceptions. For example, in question 5, 70% of instructors and 60% of graduate students selected answer choices A, C and E which do not include the force that the channel exerts on the ball; however neither of these choices was selected by 19% or more introductory students (see Table A1). Similarly, in question 11, 60% of instructors and 40% of graduate students selected choices A, B and E which do not include the normal force; however, these answer choices combined were only selected by 17% of introductory students (see Table A1). Moreover, in question 28, it is very interesting that almost half of both instructors (45%) and graduate students (44%) claimed that the most common incorrect answer choice selected by introductory students in the post-test is choice E, which states that no forces act on the ball because it is at rest (see Table A1). On the other hand, this answer choice was selected by only 1% of introductory students. Furthermore, 24% of instructors and 20% of graduate students selected choice A, which only included the force of gravity, an answer choice selected by only 2% of introductory students. Thus, instructors and graduate students did not identify introductory students' alternate conceptions related to identification of distinct forces in different contexts very well.

**TABLE 6.** Student alternate conceptions related to identifying forces, questions in which these alternate conceptions arise (FCI item #), percentage of introductory students who answer the questions incorrectly in the pre-test (% overall incorrect pre) and in the post-test (% overall incorrect post), incorrect answer choices on each question which uncover these alternate conceptions (incorrect answer choices), percentage of introductory students who hold the alternate conceptions based on their selection of these answer choices in the pre-test (Intro stud. alt. pre) and in the post-test (Intro stud. alt. post.) and percentage of instructors (Ins.) and graduate students (GS) who identify them as the most common incorrect answer choices (Ins.). For convenience, brief descriptions of the problems are given underneath.

| Introductory student alternate conceptions | FCI item # | % overall incorrect pre | % overall incorrect post | Incorrect answer choices | Intro stud. alt pre | Intro stud. alt. post | Ins. | GS |
|---|---|---|---|---|---|---|---|---|
| Do not know about any forces (including the force of gravity) | 11 | 86% | 65% | E | 3% | 4% | 20% | 0% |
| | 29 | 58% | 29% | E | 4% | 1% | 45% | 44% |
| Do not know about contact forces (normal force, tension) | 5 | 90% | 76% | A, C, E | 36% | 15% | 47% | 32% |
| | 11 | 86% | 65% | A, B, E | 41% | 17% | 60% | 40% |
| | 18 | 88% | 72% | A, C, E | 62% | 30% | 70% | 48% |
| | 29 | 58% | 29% | A, E | 19% | 3% | 69% | 65% |
| Moving objects are acted on by a distinct force in the direction of motion | 5 | 90% | 76% | C, D, E | 86% | 73% | 80% | 100% |
| | 11 | 86% | 65% | B, C | 76% | 56% | 63% | 80% |
| | 18 | 88% | 72% | C, D, E | 86% | 71% | 84% | 96% |
| **Questions** | | | | | | | | |
| 5. Identify the forces acting on a ball while moving in a frictionless, circular channel. | | | | | | | | |
| 11. Identify the forces acting on a puck while moving on a frictionless surface. | | | | | | | | |
| 18. Identify the forces acting on a boy while swinging on a rope. | | | | | | | | |
| 29. Identify the forces acting on a chair at rest on a floor. | | | | | | | | |

**I. d. iii) Constant force implies constant velocity: This alternate conception of introductory students and the performance of both instructors and graduate students in identifying it are both context dependent.**

Examination of Table 7 reveals that this alternate conception in introductory students' responses to different questions arises more or less often depending on the context. Table 7 shows that, on FCI questions 17 and 26, the vast majority of students (72% and 73%) select answer choices which imply that a constant net force would cause a constant velocity. In question 25, about half answered that the force exerted by the woman has to be greater than the total force which resists the motion of the box in order for the box to move at a constant velocity. In questions 21 and 22, the fraction of students who selected answer choices corresponding to this alternate conception was about one third and in question 24, the fraction was about one fifth. Thus, this alternate conception is observed in introductory students' responses more or less frequently depending on the context.

**TABLE 7.** Alternate conception that constant net force implies constant velocity, questions in which this alternate conception arises (FCI item #), percentage of introductory students who answer the questions incorrectly in the pre-test (% overall incorrect pre) and in the post-test (% overall incorrect post), incorrect answer choices on each question which uncovers this alternate conception (incorrect answer choices), percentage of introductory students who hold the alternate conception based on their selection of these answer choices in the pre-test (Intro stud. alt. pre) and in the post-test (Intro stud. alt. post.) and percentage of instructors (Ins.) and graduate students (GS) who identify them as the most common incorrect answer choices (Ins.). For convenience, brief descriptions of the problems are given underneath.

| Introductory student alternate conception | FCI item # | % overall incorrect pre | % overall incorrect post | Incorrect answer choices | Intro stud. alt pre | Intro stud. alt. post | Ins. | GS |
|---|---|---|---|---|---|---|---|---|
| Constant net force implies constant velocity (also: zero net force implies decreasing velocity) | 17 | 92% | 76% | A, D | 82% | 72% | 90% | 88% |
| | 21 | 65% | 67% | C | 23% | 38% | 43% | 44% |
| | 22 | 70% | 55% | A | 37% | 33% | 67% | 28% |
| | 24 | 37% | 30% | C | 25% | 22% | 70% | 68% |
| | 25 | 88% | 77% | D | 58% | 53% | 57% | 44% |
| | 26 | 97% | 86% | A, B | 83% | 73% | 87% | 74% |
| | 27 | 46% | 42% | A | 31% | 26% | 63% | 68% |
| **Questions** | | | | | | | | |
| 17. Elevator being pulled up by a cable at constant speed. | | | | | | | | |
| 21. Rocket drifting horizontally, constant thrust applied vertically, Find path followed by the rocket. | | | | | | | | |
| 22. What is the speed of the rocket during this time (constant, increasing etc.)? | | | | | | | | |
| 24. What is the speed of the rocket after thrust drops to zero (constant, increasing etc.)? | | | | | | | | |
| 25. Constant horizontal force exerted on a box which causes it to move at constant speed. | | | | | | | | |
| 26. Force in question 25 is doubled, what happens to speed of box? | | | | | | | | |
| 27. Force is removed. The box will (A) immediately come to stop, (B) continue moving at constant speed for a while and then slow to a stop, etc. | | | | | | | | |

Table 7 suggests that the performance of both instructors and graduate students in identifying this alternate conception, constant force implies constant velocity, is also context dependent and their performance varies significantly depending on the question. For example, the contexts of problems 17 and 25 are similar and in both cases an object is acted upon by two forces, one of which is applied in the direction of motion, and the other opposite to it. In question 17, they are the force exerted by the cable and the weight of the elevator and in question 25 they are the force exerted by the woman and the total force which resists the motion of the box. However, the performance of instructors and graduate students at identifying the alternate conception in these two questions is very different. In particular, in question 17 nearly all of them identified it (90% of instructors and 88% of graduate students – see Table 7) whereas in question 25, 57% of instructors and 44% of graduate students identified it. The rest of their choices regarding the most common incorrect answer of introductory students were spread over answer choices A, B and E, none of which was selected by more than 12% of introductory students (see Table A1).

In addition, the performance of instructors and graduate students related to the "constant force implies constant velocity" alternate conception is not only context-dependent, but it is also not well correlated with the strength of the alternate conception. While instructors and graduate students performed well in the two questions in which more than 70% of introductory students selected answer choices which revealed this alternate conception, they performed better in the question with a medium level alternate conception (question 24) than in other questions in which this alternate conception was strong. In particular, in question 21 and question 25, many instructors and graduate students, and in question 22, many graduate students had difficulty in identifying this alternate conception as shown in Table 7).

We note that there is a large discrepancy between the performance of instructors and graduate students in question 22. While the majority of instructors (67%) correctly identified the alternate conception that constant net force implies constant velocity on this question, fewer graduate students identified it (28%) and a large percentage of them (40%) thought that the most common alternate conception is that the speed of the rocket would increase for a while and be constant thereafter, an answer choice (choice D) selected by fewer introductory students (see Table A1). In addition, 24% of the graduate students mistakenly selected the correct answer choice as the most common incorrect answer chosen by the introductory students for this question.

**I. d. iv) Confusion between position and velocity and velocity and acceleration: Graduate students are better than instructors at identifying that some introductory students confuse position with velocity, while instructors are somewhat better than graduate students at identifying that some introductory students confuse velocity with acceleration.**

Table 8 shows that in question 19, graduate students performed better than instructors at identifying that the most common difficulty of introductory students is confusion between position and velocity. In particular, they selected answer choice D much more frequently than instructors (76% compared to 38%). Answer choice D states that the instances when the two blocks have the same speed are when the two blocks have identical positions. The answers of the instructors were spread over other answer choices which were selected by 12% or fewer introductory students (see Table A1).

In question 20, instructors performed better (although not significantly so) at identifying that the most common difficulty of introductory students is confusion between velocity and acceleration. In particular, more instructors selected answer choice D compared to the graduate students (72% instructors compared to 56% graduate students as shown in Table 8). Answer choice D states that the acceleration of block "b" is greater than the acceleration of block "a", while the strobe diagram implies that the velocity of block "b" is greater than the velocity of block "a" (both velocities are constant).

We note that for question 19 there is virtually no change in the performance of algebra-based students from the pre-test to the post-test (54% in the pre-test, 51% in the post-test). There was an improvement in the performance of introductory students in question 20 (17% improvement from 32% correct to 49% correct). One reason why it is more difficult for students to improve in performance in question 19 compared to question 20 is due to the fact that in question 19, one motion is accelerated whereas for question 20, both blocks move at a constant velocity [1-3].

**TABLE 8.** Student difficulties with interpreting strobe diagrams of motion, questions in which these difficulties arise (FCI item #), percentage of introductory students who answer the questions incorrectly in the pre-test (% overall incorrect pre) and in the post-test (% overall incorrect post), incorrect answer choices on each question which uncover the difficulties (incorrect answer choices), percentage of introductory students who have these difficulties based on their selection of these answer choices in the pre-test (Intro stud. alt. pre) and in the post-test (Intro stud. alt. post.) and percentage of instructors (Ins.) and graduate students (GS) who identify them as the most common incorrect answer choices (Ins.). For convenience, brief descriptions of the problems are given underneath.

| Introductory student difficulties | FCI item # | % overall incorrect Pre | % overall incorrect post | Incorrect answer choices | Intro stud. alt pre | Intro stud. alt. post | Ins. | GS |
|---|---|---|---|---|---|---|---|---|
| Confusing position with velocity | 19 | 46% | 49% | D | 26% | 29% | 38% | 76% |
| Confusing velocity with acceleration | 20 | 68% | 51% | C | 36% | 27% | 72% | 56% |
| **Questions** | | | | | | | | |
| 19. Diagrams of positions of two blocks at regular, successive time intervals. One block is accelerating, the other has constant velocity. Do they ever have the same speed? | | | | | | | | |
| 20. Diagrams of positions of two blocks at regular, successive time intervals. Both blocks move at constant velocities, one smaller than the other. Compare the accelerations. | | | | | | | | |

## I. d. v) Instructors' and graduate students' difficulties in identifying other common alternate conceptions of introductory students

The student alternate conception related to an impetus view of motion identified in question 13 is that after a boy throws a ball in the air vertically, on the way up, in addition to the force of gravity, a steadily decreasing force also acts on the ball. On the way down, only the force of gravity acts on the ball. This alternate conception (which is held by 50% of introductory students) was identified by about half of the instructors (47%), but very few graduate students (16%) as shown in Table 10. A sizeable percentage of both instructors (30%) and graduate students (44%) thought that the most common incorrect answer choice of introductory students for question 13 is choice B (see Table A1) in which on the way down, the force of gravity steadily increases. Only 11% of introductory students selected this answer choice. In addition, 20% of instructors and 36% of graduate students selected answer choice A as the most common alternate conception, which does not make a distinction between the forces acting on the object on the way up and on the way down (downward force of gravity along with a steadily decreasing upward force), and which was selected by only 4% of introductory students. Thus, the responses of instructors and graduate students suggest that they do not have a good understanding of introductory physics students' difficulty in this situation.

Question 14 reveals an interesting introductory student difficulty. Although the question was somewhat easy for introductory students in the post-test (61% correct), 19% of introductory students selected the trajectory which arches backwards even in the post-test. This question is one for which the reasons for explicitly selecting the answers are not provided, and it would be worthwhile knowing students' reasoning. We therefore added reasons for the each answer choice and administered the question as part of a final exam in a large algebra-based introductory physics class with 400+ students. The reasons for the path that arches backwards were (A) "because by the time it strikes the ground, the plane will cover some horizontal distance" and (B) "due to air resistance". Choice (C) in this multiple-choice question provided a justification for path (2) which goes straight down: "the force of gravity is the only force acting on the ball after the ball falls from the plane and it causes the ball to fall vertically downwards". These justifications increased the percentage of students who selected these answer choices by 5% each. The percentages of students who selected each incorrect answer choice (A), (B) and (C) are: 17%, 7% and 14%. It appears that the main reason students select the path that arches backwards is because they are having difficulty viewing the motion of the ball from the perspective of a person on the ground. They are implicitly in the airplane thinking that it keeps travelling after the bowling ball falls out and therefore covers more horizontal distance. In the FCI version of the question, few instructors (17%) and graduate students (20%) identified that the most common incorrect choice would be (A), the path the arches backwards. Both groups selected choices B (straight down) and C (straight oblique line) much more often (see Table A1), and these answer choices were selected by fewer introductory students (10% and 9% – see Table A1).

**TABLE 10.** Three other common alternate conceptions/difficulties, questions in which these difficulties arise (FCI item #), percentage of introductory students who answer the questions incorrectly in the pre-test (% overall incorrect pre) and in the post-test (% overall incorrect post), incorrect answer choices on each question which uncover the difficulties (incorrect answer choices), percentage of introductory students who have these difficulties based on their selection of these answer choices in the pre-test (Intro stud. alt. pre) and in the post-test (Intro stud. alt. post.) and percentage of instructors (Ins.) and graduate students (GS) who identify them as the most common incorrect answer choices (Ins.). For convenience, brief descriptions of the problems are given underneath.

| Introductory student alternate conceptions/difficulties | FCI item # | % overall incorrect pre | % overall incorrect post | Incorrect answer choices | Intro stud. alt pre | Intro stud. alt. post | Ins. | GS |
|---|---|---|---|---|---|---|---|---|
| Ball thrown vertically in the air: on the way up - steadily decreasing upward force and gravity, on way down, only gravity | 13 | 88% | 65% | C | 64% | 50% | 47% | 16% |
| Relative velocity and reference frame difficulties | 14 | 64% | 39% | A | 35% | 19% | 17% | 20% |
| If a constant force acts on an object for some time and then it is removed, the object will eventually go back to the direction in which it was originally moving | 23 | 71% | 61% | D | 28% | 23% | 23% | 24% |
| **Questions** | | | | | | | | |
| 13. Ball thrown vertically in the air, no air resistance. Find the forces acting on the ball while in the air. | | | | | | | | |
| 14. Bowling ball rolls off a plane while plane is travelling horizontally. Find the path of the ball. | | | | | | | | |
| 23. Rocket moving horizontally. Constant thrust applied vertically for some time then removed. Find the path of the rocket after thrust drops to zero. | | | | | | | | |

Question 23 reveals another interesting student alternate conception. For this question, the most commonly selected answer choice (by 23% of introductory students) is choice D. The path described by choice D is one in which the direction of the velocity of the rocket gradually returns to its original orientation (to the right). This implies that students who selected this choice thought that forces which act for a finite time do not change the direction of motion indefinitely and the rocket eventually returns to its original orientation. Only about one quarter of both instructors and graduate students (23% and 24% – see Table 10) identified this as the most common incorrect answer choice. The answers of graduate students appear to suggest that they are random guessing (percentages between 20% and 28% for each incorrect answer choice – see Table A1), while 47% of instructors selected answer choice C (vertical path), which was selected by only 18% of introductory students. Thus, in this context also, the instructors and graduate students struggled to identify the most common difficulties of introductory physics students.

### II. Introductory student FCI performance

### II. a. Which questions on the FCI pose significant challenges for students even after instruction (poor performance)?

We note that the original paper describing the development of the FCI has a discussion of students' difficulties in the original version of the FCI even after instruction. For the later version of the FCI that we employed in our research, this question about introductory physics students' difficulties in post-test was answered earlier while discussing which introductory student alternate conceptions were correctly identified by instructors and graduate students. One can also refer back to Table A1 which provides the performance of introductory students on the post-test on the questions in the FCI in which at least one incorrect answer choice was selected by 19% or more introductory students.

### II. b. Are there any questions on the FCI in which there is little improvement (less than 10%) for algebra-based students from pre- to post-test?

The researchers decided that "little" (not noteworthy) improvement occurred from pre- to post-test in questions in which the normalized gain was less than 0.173 (i.e. normalized gain is in the lower 1/3 based on the average normalized gain of 0.26), and questions in which the percentage of introductory students who hold a particular alternate conception decreased by 5% or less. There were twelve questions on the FCI (shown in Table 11) which fit at least one of these two criteria. Table 11 also shows the percentage of introductory students who answered each question incorrectly both in the pre-test and in the post-test, the normalized gain, the incorrect answer choices corresponding to the most common alternate conceptions and the percentage of introductory students who hold those alternate conceptions both in the pre-test and in the post-test.

**TABLE 11.** The 12 questions on the FCI on which there was little improvement (less than 0.173 normalized gain or difference of 5% or less in the percentages of introductory physics students harboring a particular alternate conception), student alternate conceptions/difficulties associated with these questions, percentage of introductory students who answered them incorrectly in the pre-test (% incorrect pre) and in the post-test (% incorrect post), normalized gain (Norm. gain), most common incorrect answer choices which uncovered these alternate conceptions/difficulties (incorrect answer choices), percentage of students who have these alternate conceptions/difficulties based on their selection of those incorrect answer choices in the pre-test (Intro stud. alt. pre) and in the post-test (Intro stud. alt. post). For convenience, short descriptions of the questions are given underneath.

| FCI item # | Introductory student alternate conceptions/difficulties | % overall incorrect pre | % overall incorrect post | Norm. gain | Incorrect answer choices | Intro stud. alt. pre | Intro stud. alt. post |
|---|---|---|---|---|---|---|---|
| 5 | Moving objects have a distinct force in the direction of motion | 90% | 76% | 0.16 | C, D, E | 86% | 73% |
| 30 | | 88% | 74% | 0.16 | B, D, E | 87% | 71% |
| 9 | Difficulties with addition of perpendicular velocities | 57% | 47% | 0.17 | C | 20% | 19% |
| 19 | Confusing position with velocity | 46% | 49% | -0.06 | D | 26% | 29% |
| 23 | If a constant force acts on an object for some time and then it is removed, the object will eventually go back to the direction in which it was originally moving | 71% | 61% | 0.14 | D | 28% | 23% |
| 17 | | 92% | 76% | 0.17 | A | 60% | 62% |
| 21 | | 65% | 67% | -0.02 | C | 23% | 38% |
| 22 | Constant net force implies constant velocity (also: zero net force implies decreasing velocity) | 70% | 55% | 0.22 | A | 37% | 33% |
| 24 | | 37% | 30% | 0.20 | C | 25% | 22% |
| 25 | | 88% | 67% | 0.13 | D | 58% | 53% |
| 26 | | 97% | 86% | 0.11 | A | 41% | 41% |
| 27 | | 46% | 42% | 0.09 | A | 31% | 26% |

**Questions**

5. Identify the forces acting on a ball while moving in a frictionless, circular channel.

9. Puck sliding horizontally with speed "$v_0$". Kicked vertically (if at rest kick would give the puck speed "$v_k$". What is the speed of the puck just after the kick?

17. Elevator being pulled up by a cable at constant speed.

19. Diagrams of positions of two blocks at regular, successive time intervals. One block is accelerating, the other has constant velocity. Do they ever have the same speed?

21. Rocket drifting horizontally, constant thrust applied vertically. Find path followed by the rocket.

22. What is the speed of the rocket during this time (constant, increasing, etc.)?

23. Path of the rocket after thrust drops to zero.

24. What is the speed of the rocket after thrust drops to zero (constant, increasing, etc.)?

25. Constant horizontal force exerted on a box which causes it to move at constant speed.

26. Force in question 25 is doubled, what happens to speed of box?

27. Woman stops applying horizontal force (from question 26). The box will: immediately come to a stop, immediately start slowing to a stop etc.

30. Tennis player hits a tennis ball against strong wind. Identify the forces acting on the tennis ball while in the air.

**Constant net force implies constant velocity and zero net force implies decreasing velocity**

The most prevalent difficulty observed was that introductory students have a very difficult time abandoning the notion that a constant net force implies a constant velocity. In the questions which can be used to test for this alternate conception (questions 17, 21, 22, 24, 25, 26 and 27), there was either less than 0.173 normalized gain, and/or the percentage of students who hold this alternate conception did not decrease by more than 5% (see Table 11). In question 21, there was a slight shift in alternate conceptions because in the pre-test, students selected answer choice B (vertical path, i.e., ignored the initial motion of the rocket) and choice C (oblique straight path, i.e., implying that constant vertical force results in constant vertical speed) equally (21% and 23%), however, in the post-test, more students selected choice C than choice B (38% compared to 13%). It appears that some students have learned that the initial motion of the rocket must be taken into account when determining the path after the engine of the rocket was turned on, but they still harbor the alternate conception that a constant force implies a constant velocity.

**Moving objects are acted upon by a distinct force in the direction of motion**

When it comes to this alternate conception, it appears that introductory students improved in some contexts after instruction, but not in others. Out of the four questions in which this alternate conception can be identified, the normalized gain was less than 0.173 in two (questions 5 and 30 – see Table 11). However, even in the other two questions in which the normalized gain was not in the lower one third, it was not very large (question 11 – puck sliding across a frictionless surface: normalized gain = 0.24 and question 18 – boy swinging on a rope – normalized gain = 0.19 – see Table A3).

**Difficulties with addition of perpendicular velocities**

In question 9, it appears that the same number of students (20% in the pre-test and 19% in the post-test) noted that the final speed of the puck will be the arithmetic sum of "$v_0$" (initial speed of the puck) and "$v_k$" (the speed the kick would have imparted, had the puck been stationary). These students had difficulty realizing that the Pythagorean Theorem must be applied to add the two perpendicular velocities. Even after instruction, only about half of the introductory students correctly reasoned that the final speed will be greater than either of the speeds "$v_0$" and "$v_k$", but less than their sum.

**Confusing position with velocity**

In question 19, which assesses students' ability to extract information about speed from strobe diagrams of motion, the percentage of correct answers decreased from 54% before instruction to 51% after instruction. In addition, the percentage of introductory students who confused position with velocity (choice D: the two objects have the same speed at points 2 and 5 on the strobe diagram at which points they have the same position) remains approximately the same (26% in the pre-test and 29% in the post-test). Interestingly, in question 20, which assesses students' ability to extract information about acceleration from strobe diagrams of motion, the normalized gain was 0.25 (see Table A3).

**If a constant force acts on an object for some time and then it is removed, the object will eventually return to the direction in which it was originally moving before the constant force was applied**

Many introductory physics students incorrectly believe this**.** In addition to low normalized gain on question 23, which assessed understanding of this concept (normalized gain = 0.11 – see Table 11), the percentage of students who hold this alternate conception decreased by only 5% from the pre-test to the post-test.

### II. c. Are there any shifts in the most common alternate conceptions from the pre-test to the post-test?

The only major shift in alternate conceptions of introductory students which occurred for more than one question was observed on questions which asked students to identify all the distinct forces that act on an object. Before instruction, algebra-based students selected incorrect answer choices which corresponded to the force of gravity and force in the direction of motion with similar frequency compared to the incorrect answer choices which corresponded to the force of gravity, contact forces and force in the direction of motion. After instruction, they overwhelmingly selected the latter compared to the former. The other shift was only in question 2, for which, before instruction, more algebra-based students thought that the ball twice as heavy will strike the floor considerably closer compared to students who thought that the ball twice as heavy will strike the floor at exactly half the distance of the lighter ball, whereas after instruction the percentages of students who held these alternate conceptions are about the same. Table 12 shows these alternate conceptions, the questions in which they occur, the percentage of incorrect answers both in the pre-test and in the post-test along with the incorrect answer choices corresponding to the most common alternate conceptions and the percentage of students who hold these alternate conceptions.

**Identify all of the distinct forces that act on an object**

For questions 5, 11 and 18, the answer choices which include as distinct forces the force of gravity and a force in the direction of motion are choices C, B and C, respectively, while the answer choices which include as distinct forces the force of gravity, the contact force and a force in the direction of motion are choices D, C and D, respectively. Table 12 shows that the percentage of introductory students who selected these incorrect answer choices in each question are comparable in the pre-test (question 5: C – 31%, D – 25%; question 11: B – 31%, C – 45% and question 18: C – 14% and D – 27%). In the post-test, the incorrect answer choices shift significantly towards the answer choices which include the force of gravity, the contact force and a force in the direction of motion (question 5: 44% compared to 12%, question 11: 48% compared to 8%, question 18: 42% compared to 4%). It appears that before instruction, some students are not aware of contact forces (normal force, tension force), and after instruction they are aware of them. However, introductory students often do not abandon the alternate conception that if an object is moving in a certain direction, a distinct force must be acting on it in that direction.

**TABLE 12.** Introductory students' alternate conceptions related to identifying all the distinct forces that act on an object, and alternate conceptions related to question 2, the questions in which these alternate conceptions occurred, the percentage of introductory students who answered the questions incorrectly in the pre-test (% incorrect pre) and in the post-test (% incorrect post), the most common incorrect answer choices which uncovered these alternate conceptions and the percentage of students who hold these alternate conceptions based on their selection of those incorrect answer choices in the pre-test (Intro stud. alt. pre) and in the post-test (Intro stud. alt. post). For convenience, short descriptions of the questions are given underneath.

| Introductory student alternate conceptions | FCI item # | % overall incorrect pre | % overall incorrect post | Incorrect answer choices | Intro stud. alt. pre | Intro stud. alt. post |
|---|---|---|---|---|---|---|
| Force of gravity and force in the direction of motion | 5 | 90% | 76% | C | 31% | 12% |
| | 11 | 86% | 65% | B | 31% | 8% |
| | 18 | 88% | 72% | C | 14% | 4% |
| Force of gravity, contact force and force in the direction of motion | 5 | 90% | 76% | D | 25% | 44% |
| | 11 | 86% | 65% | C | 45% | 48% |
| | 18 | 88% | 72% | D | 27% | 42% |
| Ball twice as heavy that rolls off horizontal table travels half as far | 2 | 73% | 56% | B | 21% | 25% |
| Ball twice as heavy that rolls off horizontal table travels considerably less, but not half | 2 | 73% | 56% | C | 37% | 21% |
| **Questions** | | | | | | |
| 2. Two metal balls, same size, one twice as heavy as other, roll off a horizontal table: (A) both balls hit the floor at the same distance, (B) heavier ball hits the floor at half the distance of the lighter ball, etc. | | | | | | |
| 5. Identify the forces acting on a ball while moving in a frictionless, circular channel. | | | | | | |
| 11. Identify the forces acting on a puck while moving on a frictionless surface. | | | | | | |
| 18. Identify the forces acting on a boy while swinging on a rope. | | | | | | |

**Two metal balls roll off a horizontal table**

The other shift in alternate conceptions occurred in question 2. In the pre-test, more students thought that the heavier ball hits the floor considerably closer than the lighter ball, but not necessarily half the horizontal distance, compared to students who thought that it hits the floor at half the distance of the lighter ball (37% compared to 21% – see Table 12). In the post-test however, the percentages of students who selected these choices are about the same (21% and 25%).

## II. c. On which questions do calculus-based students perform better than algebra-based students? Are there any questions in which the alternate conceptions of algebra-based students are different from the alternate conceptions of calculus-based students?

Due to the large population sizes, any difference of 5% or more turned out to be statistically significant by means of chi-square tests [55]. However, a difference of 5% in performance from the pre-test to the post-test does not have much practical significance. Instead, questions which were answered correctly by 20% or more of calculus-based students compared to algebra-based students were chosen as a heuristic by the researchers to be indicative of significantly better performance of calculus-based students compared to algebra-based students. The question about whether the alternate conceptions of algebra-based students are different from the alternate conceptions of calculus-based students was answered by investigating whether there were any questions in which the most common incorrect answer choice(s) of algebra-based students was (were) different from the most common incorrect answer choice(s) of calculus-calculus based students. It turned out that there were no such questions. For all questions which included more than one common incorrect answer choice, this was most common for both algebra-based and calculus-based students (in addition, the fraction of calculus-based students who selected that particular incorrect answer choice was always smaller than the fraction of algebra-based students – see Tables A3 and A4). Similarly, for the questions which included two common incorrect answer choices, they were common for both algebra-based and calculus-based students (see Tables A3 and A4). Moreover, only one question had three common incorrect answer choices and these three answer choices were the most common incorrect answers for both algebra-based and calculus-based students. It therefore appears that in the pre-test, the algebra-based students harbor the same alternate conceptions as calculus-based students. However, algebra-based students hold the same alternate conceptions more strongly than calculus-based students.

### II. c. i) Pre-test comparison of performance of algebra-based and calculus-based students

Calculus-based students correctly answered every single question on the FCI more frequently than algebra-based students in the pre-test. Differences of 10% or more occurred on 26 questions and differences of 20% or more occurred on 8 questions (see Tables A3 and A4). We will focus on the questions in which the differences were of 20% or more (questions 1, 3, 12, 13, 14, 20, 22 and 28).

**Significantly better performance of calculus-based students compared to algebra-based students is context dependent**

An interesting finding suggested by Table 13 is that calculus-based students answer questions involving particular force and motion concepts significantly better than algebra-based students (by 20% or more) in some contexts, but not in others. For example, 20% more calculus-based students than algebra-based students correctly interpreted Newton's third law in the context of problem 28 (one student pushing another). However, in the other two questions with the alternate conception that the active object exerts more force on the passive object than vice versa (question 15 – car pushing truck and accelerating and question 16 – car pushing truck at constant speed) calculus-based students did not outperform algebra-based students by more than 20%. In fact, the difference in question 15 is merely 2% (see Tables A3 and A4).

A similar observation can be made by examining the questions related to the alternate conception that a constant net force implies a constant velocity (questions 17, 21, 22, 24, 25, 26 and 27). We find that 21% more calculus-based students than algebra-based students answered question 22 correctly. However, in the other questions involving the same concept, the smallest difference in the performance of calculus-based and algebra-based students was 11% (i.e., calculus-based students always performed better, but not always by 20% or more).

**TABLE 13.** Questions in which calculus-based students outperformed algebra-based students in the pre-test, the most common alternate conceptions/difficulties uncovered by these questions, percentage of incorrect answers for both algebra-based (% overall incorrect algebra) and calculus-based (% overall incorrect calculus) introductory students, incorrect answer choices which correspond to the most common alternate conceptions/difficulties (incorrect answer choices) and percentages of algebra-based (Alg. alt.) and calculus-based (Calc. alt.) students who harbor/have these alternate conceptions/difficulties

| FCI item # | Pre-test introductory student alternate conceptions | % overall incorrect algebra | % overall incorrect calculus | Incorrect answer choices | Alg. alt. | Calc. alt. |
|---|---|---|---|---|---|---|
| 1 | Time it takes an object to fall freely through a certain distance is proportional to mass | 47% | 18% | C | 25% | 8% |
| 3 | Freely falling objects reach terminal velocity a short time after release | 60% | 34% | A | 31% | 15% |
| 12 | An object fired horizontally will not immediately descend and continue to move horizontally for some time | 41% | 16% | C | 32% | 19% |
| 13 | Ball thrown vertically in the air: on the way up - steadily decreasing upward force and gravity, on way down, only gravity | 88% | 67% | C | 64% | 50% |
| 14 | Relative velocity and reference frame difficulties | 64% | 37% | A | 35% | 19% |
| 20 | Confusing velocity with acceleration | 68% | 46% | C | 36% | 22% |
| 22 | Constant net force implies constant velocity | 70% | 49% | A | 37% | 27% |
| 28 | Newton's third law: the active object exerts more force on the passive than vice versa | 76% | 56% | D | 61% | 45% |
| **Questions** | | | | | | |
| 1. Two metal balls, same size, one twice as heavy as the other are dropped from the same height. The time it takes the balls to fall is (A) half as long for heavier ball, (B) half as long for lighter ball (C) same, etc. | | | | | | |
| 3. The two balls from question 1 roll off a horizontal table. (A) distance same for both balls, (B) distance of heavier ball is half the distance of lighter ball, etc. | | | | | | |
| 12. Ball fired horizontally from cannon. Determine the path it follows. | | | | | | |
| 13. Ball thrown vertically in the air, no air resistance. Find the forces acting on the ball while in the air. | | | | | | |
| 14. Bowling ball rolls off a plane while plane is travelling horizontally. Find the path of the ball. | | | | | | |
| 20. Diagrams of positions of two blocks at regular, successive time intervals. Both blocks move at constant velocities, one smaller than the other. Compare the accelerations. | | | | | | |
| 22. Rocket drifting horizontally, constant thrust applied vertically. Speed of rocket during this time (constant, increasing etc.) | | | | | | |
| 28. Student "a" puts his feet on student "b" and pushes against student "b". | | | | | | |

**Understanding of freely falling objects**

The better performance of calculus-based students compared to algebra-based students in questions 1 and 3 (by 29% and 26% respectively) in the pre-test indicates that calculus-based students have a better understanding of the physics of freely falling objects before instruction.

**An object fired horizontally will not immediately descend, but continue to move horizontally for some time**

The data in Table 13 suggest that algebra-based students performed worse than calculus-based students on question 12, which uncovered this alternate conception, by 25%.

**Relative velocity and reference frame difficulties**

In question 14, it appears that algebra-based students find it difficult to view the motion of the bowling ball falling from the airplane from the correct frame of reference. Many introductory students thought that the path of the ball falling from the plane arches backwards because they have difficulty viewing the path of the ball as ground observers [1-3].

**Impetus view of motion**

Question 13 indicates that many algebra-based (64%) and calculus-based (50%) students have an impetus view of motion before instruction (ball thrown vertically in the air, on the way up will be acted upon by a steadily decreasing upward force and the force of gravity, and on the way down, will only be acted upon by the force of gravity), however, 21% more calculus-based than algebra-based students answer this question correctly.

**Confusion between velocity and acceleration**

Question 20, which was answered correctly by 22% more calculus-based students than algebra-based students, indicates that calculus-based students are more likely to correctly interpret acceleration from strobe diagrams of motion. The most common incorrect answer choice for both groups is choice C (acceleration of "b" is greater than that of "a") which indicates that many introductory students confuse acceleration with velocity of an object.

## II. c. iii) Post-test comparison of performance between algebra-based and calculus-based students

Similar to the pre-test, calculus-based students outperformed algebra based students on all but one question (item 15) after instruction (post-test). Differences of 10% or more occurred on 26 questions and differences of 20% or more occurred on 14 questions (more than in the pre-test for which differences of 20% or more occurred on only 8 questions). These questions are 5, 9, 10, 11, 13, 17 through 23, 25 and 26. Question 10 is not included in Table 12 because although calculus-based students performed better than algebra-based students by 20%, there were no incorrect answer choices selected by 19% or more of either calculus-based or algebra-based students in the post-test, and therefore no strong or medium level common alternate conceptions were uncovered by this question.

**Identifying all of the distinct forces that act on an object**

Questions 5, 11 and 18 all ask students to identify all of the distinct forces acting on an object. Comparison of algebra-based students' alternate conception shifts in these questions indicated that on the pre-test many of them had failed to identify contact forces, while on the post-test, they do identify them, however, they retain the alternate conception that moving objects are acted upon by a distinct force in the direction of motion. Comparison of performance of algebra-based students with calculus-based students for the post-test indicates that more algebra-based than calculus-based students, even after instruction, still claim that if an object is **TABLE 14.**

Questions on which calculus-based students outperformed algebra based students in the post-test, the most common alternate conceptions/difficulties uncovered by these questions, percentage of incorrect answers for both algebra-based (% overall incorrect algebra) and calculus-based (% overall incorrect calculus) introductory students, incorrect answer choices which correspond to the most common alternate conceptions/difficulties (incorrect answer choices) and percentages of algebra-based (Alg. alt.) and calculus-based (Calc. alt.) students who have these alternate conceptions/difficulties.

| FCI item # | Post-test introductory student alternate conceptions/difficulties | % overall incorrect algebra | % overall incorrect calculus | Incorrect answer choices | Alg. alt. | Calc. alt. |
|---|---|---|---|---|---|---|
| 5 | Moving objects have a distinct force in the direction of motion | 76% | 53% | C, D | 56% | 42% |
| 11 | | 65% | 39% | B, C | 56% | 30% |
| 18 | | 72% | 45% | C, D, E | 71% | 46% |
| 17 | Constant net force implies constant velocity (also: zero net force implies decreasing velocity) | 76% | 56% | A, D | 72% | 52% |
| 21 | | 67% | 43% | C | 38% | 23% |
| 22 | | 55% | 33% | A | 33% | 20% |
| 25 | | 77% | 49% | D | 53% | 38% |
| 26 | | 86% | 58% | A, B | 73% | 58% |
| 9 | After performing an action on an object, its speed depends only on the action, not the previous motion | 47% | 27% | B, C | 39% | 23% |
| 13 | Ball thrown vertically in the air: on the way up - steadily decreasing upward force and gravity, on way down, only gravity | 65% | 39% | C | 50% | 31% |
| 19 | Confusing position with velocity | 49% | 25% | D | 29% | 12% |
| 20 | Confusing velocity with acceleration | 51% | 29% | C | 38% | 23% |
| 23 | If a constant force acts on an object for some time and then it is removed, the object will eventually return to the direction in which it was originally moving | 61% | 36% | D | 23% | 14% |

5. Identify the forces acting on a ball while moving in a frictionless, circular channel.

9. Puck sliding horizontally with speed "$v_0$". Kicked vertically (if at rest kick would give the puck speed "$v_k$". What is the speed of the puck just after the kick?

11. Identify the forces acting on a puck while moving on a frictionless surface.

13. Ball thrown vertically in the air, no air resistance. Find the forces acting on the ball while in the air.

17. Elevator being pulled up by a cable at constant speed.

18. Identify the forces acting on a boy while swinging on a rope.

19. Diagrams of positions of two blocks at regular, successive time intervals. One block is accelerating, the other has constant velocity. Do they ever have the same speed?

20. Diagrams of positions of two blocks at regular, successive time intervals. Both blocks move at constant velocities, one smaller than the other. Compare the accelerations.

21. Rocket drifting horizontally, constant thrust applied vertically, find path followed by the rocket.

22. Speed of the rocket during this time (constant, increasing, etc.)?

23. Rocket moving horizontally. Constant thrust applied vertically for some time, then removed. Find the path of the rocket after thrust is removed.

25. Constant horizontal force exerted on a box which causes it to move at constant speed.

26. Force in question 25 is doubled, what happens to speed of box?

moving in a certain direction, a distinct force must be acting on the object in the same direction. Question 30 is similar, and in this question as well, more algebra-based students than calculus-based students think that there is a force of the "hit" that continues to act on the ball even when the tennis ball loses contact with the racquet. In particular, the calculus-based students outperformed the algebra-based students by 18% on this question.

**Constant net force implies constant velocity and zero net force implies decreasing velocity**

Calculus-based students outperformed algebra-based students by at least 20% in almost all questions on the FCI in which this alternate conception is uncovered (see Table 14). Furthermore, the largest discrepancies between students in the calculus-based and algebra-based courses on all FCI questions occurred in questions 25 and 26 (28%). It appears that calculus-based students are better than algebra-based students at discarding the alternate conception that constant net force implies constant velocity and improving their performance in questions dealing with Newton's $2^{nd}$ law. In particular, on the pre-test, calculus-based students outperformed algebra-based students on only one question (question 22), which dealt with the alternate conception that constant net force implies constant velocity, but in the post-test, on all these questions, they improved more than algebra-based students, both in the percentage of correct answers and in the percentage of students who hold this alternate conception (see Tables A3 and A4).

**Impetus view of motion**

On question 13, the performance of calculus-based students was better in the pre-test (by 21%) as well. The discrepancy in performance is slightly higher in the post-test (26%).

**After applying a force on an object, its speed depends only on the applied force, and not on the previous motion**

On question 9, more algebra-based students retained the alternate conceptions that the speed of the puck after receiving the kick would be the same as the speed the kick would impart if the puck was stationary and independent of the original speed of the puck. Calculus-based students performed better than algebra-based students on this question by 20% (see Table 14).

**Interpreting strobe diagrams of motion**

Calculus-based students outperformed algebra-based students in both questions 19 and 20 which assess students' ability to extract information about velocity and acceleration from strobe diagrams of motion. In particular, algebra-based students are more likely than calculus-based students to confuse position with velocity (in question 19) and velocity with acceleration (in question 20).

**II. c. iv): Post-test comparison of alternate conceptions between algebra-based and calculus-based students**

Similar to the pre-test, in the post-test, the most common alternate conceptions were the same for algebra-based and calculus-based students. However, on almost all questions, algebra-based students held these alternate conceptions more strongly than calculus-based students (see Tables A3 and A4).

**Discussion and Summary**

Awareness of introductory physics students' difficulties and being able to understand the way they reason about physics is important because instruction can take advantage of students' initial knowledge and pedagogical approaches and curricula can explicitly account for these difficulties. Our investigation involved using the FCI to evaluate the pedagogical content knowledge of both instructors and Teaching Assistants (TAs) with varying degree of teaching experience. For each item on the FCI, the instructors and TAs were asked to identify the most common incorrect answer choice of introductory physics students. We also discussed the responses individually with a few instructors and in a class discussion with the graduate students.

**The ability to identify common introductory students' alternate conceptions in the FCI does not appear to be dependent on teaching experience or familiarity with US teaching practices**

We find that the instructors, who on the average had significantly more teaching experience as lecturers, did not perform better at identifying common introductory student alternate conceptions than graduate students, who had limited teaching experience as lecturers. We note however, that graduate students had taken introductory physics only four years prior to this study as an undergraduate student and a majority of them were TAs in an introductory recitation or laboratory class, did weekly grading of quizzes, homework and exams and held office hours in which they helped introductory students individually. These experiences may have, on average, improved their ability to identify introductory physics students' difficulties related to FCI concepts. Among both instructors and graduate students, some of them performed very well, while others performed poorly. Moreover, the ability to correctly identify students' difficulties was not correlated with the teaching experience of the physics instructors in introductory algebra-based and calculus-based mechanics courses. In particular, the performance among instructors was not better for those who had taught these courses recently (last seven years) compared to those who had not taught recently. One possible reason for why there was no statistically significant difference between the two groups of instructors is that all instructors who taught introductory mechanics employed traditional methods, most had minimal contact with students in the large introductory classes, and did not grade introductory students' homework and quizzes which may have provided some insight into students' common difficulties (the grading was done by the TAs). Moreover, even instructors in the other group who had not taught introductory mechanics had taught other introductory courses in which force concepts were relevant and many of these instructors had taught introductory mechanics more than seven years ago.

We also investigated whether the ability of American graduate students to identify introductory students' alternate conceptions was better than that of foreign graduate students and found that this was not the case. The numbers of graduate students in the different groups (American – 9, Chinese – 9 and other foreign – 7) were too small to perform meaningful statistics, but it appears that their average PCK performance is very similar. The discussions with graduate students from different countries in the TA training class about this FCI related PCK task suggested that foreign students were similar to American students in this regard, but it is difficult to justify why their PCK scores are comparable despite their different backgrounds.

**Instructors appear to identify 'strong' student alternate conceptions better than 'medium' level ones, while graduate students exhibit similar performance**

An alternate conception was considered 'strong' if it is held by more than 1/3 of introductory students. 'Medium' level alternate conceptions were connected to incorrect answer choices selected by a percentage of introductory students between 19% and 34%. We found that instructors were able to identify the strong alternate conceptions somewhat more often than the medium level ones while graduate students exhibited similar performance.

**Discussions among graduate students improved their PCK performance in identifying common student alternate conceptions**

The graduate students identified what they thought to be the most common introductory student alternate conceptions first individually and then in a group of two or three. Their group performance was statistically significantly better than their individual performance. In addition, when the individual answers of graduate students working in a group disagreed, discussions more often shifted towards the more common alternate conception (74% of the time) than on the less common one. This implies that discussing student difficulties with other TAs/instructors leads to a better understanding of students' initial knowledge state (and difficulties). Therefore, exercises which incorporate such discussions in the context of conceptual assessments could be beneficial and should be incorporated in teacher preparation and/or training courses.

**For most alternate conceptions which appear in more than one question, the ability of both instructors and graduate students to identify them is context dependent**

We find that while both physics instructors and TAs, on average, performed better than random guessing at identifying introductory students' difficulties with FCI content, they did not identify many common difficulties that introductory physics students have even after traditional instruction and their ability to identify them was context dependent. For example, for Newton's third law alternate conceptions, the vast majority of both instructors (97%) and graduate students (84%) identified the most common alternate conception in the typical context (truck colliding with car), but fewer identified it in other contexts (for example car pushing truck and accelerating – 60% of instructors and 40% of graduate students identified the most common student alternate conception that the car exerts the larger force).

Similarly, identifying the common alternate conception that a constant force implies a constant velocity was also context dependent. For example, questions 17 (elevator being pulled up by a cable at constant speed) and 25 (constant horizontal force applied on a box which causes it to move at constant speed) are similar. However, both instructors and graduate students correctly selected the alternate conception in question 25 much less frequently than in question 17 (90% compared to 57% for instructors and 88% compared to 44% for graduate students). Similar observations can be made while examining the other five questions involving this alternate conception.

For alternate conceptions related to identifying all distinct forces that act on an object, there was no context dependence in the ability of both instructors and graduate students to identify the most common student alternate conceptions; however, their PCK performance leaves a lot of room for improvement. In particular, the largest percentage of instructors who identified the most common alternate conception related to identification of distinct forces in any of these questions was 40% and for graduate students it was 60%.

**Alternate conceptions for which the PCK performance of instructors and graduate students leaves a lot of room for improvement**

As noted earlier, neither instructors nor graduate student TAs performed well at identifying student alternate conceptions related to identifying distinct forces (questions 5, 11, 18, 29 and 30). This is because in almost all these questions (all except for question 30), a sizeable majority of instructors and graduate students selected answer choices which did not include contact forces or any forces which was inconsistent with introductory student choices. In question 28, for example, (chair at rest on the floor), 44% of instructors and 45% of graduate students thought that the most common student alternate conception is that no forces act on the chair because it is at rest, an answer choice selected by only 1% of introductory students.

For introductory student difficulties related to interpreting strobe diagrams of motion, the majority of instructors did not identify that introductory students confuse position with velocity (only 38% of instructors identify this difficulty), and only half of the graduate students identify that introductory students confuse velocity with acceleration.

Alternate conceptions related to Newton's third law are identified by both instructors and graduate students in a typical context (truck colliding with car), but not in less typical contexts (questions 15, 16 and 28) for which the largest percentage of instructors who identify the most common alternate conception is 60% and for graduate students 52%. A similar observation can be made for the alternate conception that constant force implies constant velocity. However, in these questions instructors and graduate students perform reasonably well in more than half of them (5/7 for instructors and 4/7 for graduate students).

There are three other alternate conceptions/difficulties which are not identified by the majority of instructors or graduate students (for two of them, the largest percentage of instructors or graduate students who identify them is 24%). These occur on questions 13 (ball thrown vertically in the air on which students have to identify all the forces), question 14 (bowling ball rolls off a plane while the plane is moving horizontally, on which students have to determine the path of the ball as viewed from the ground) and question 23 (rocket moving horizontally, with constant thrust applied vertically for some time and then removed, on which students have to determine the path of the rocket after the thrust is removed).

In summary, there were many alternate conceptions held by more than 19% of introductory students (strong or medium level) that were not identified very often by both instructors and graduate students. Even instructors who teach introductory courses on a regular basis struggled to identify some common alternate conceptions. In addition, some instructors and graduate students explicitly noted that this task was challenging and it was difficult for them to think about physics questions from a student's perspective and expressed concern about their performance (a few noted that they were confident that they have performed poorly on this task).

**Introductory student performance – most prevalent difficulties**

The performance of introductory physics students is discussed at length in part II. b. of the results section; here we will summarize the most important results.

Introductory students have a very difficult time abandoning the alternate conception that a constant force implies constant velocity. In all the questions for which this alternate conception can surface (questions 17, 21, 22, 24, 25, 26 and 27), either the normalized gain was less than 0.175 or the percentage of students who hold this alternate conception did not decrease by more than 5% from pre-test to post-test.

The introductory students' performance on questions in which the alternate conception that moving objects are acted upon by a distinct force in the direction of motion can surface (questions 5, 11, 18 and 30) improved on some questions, but not on others. Two of the questions had normalized gains less than 0.173 and the other two had larger normalized gains, but not by much (they were 0.19 and 0.24).

The difficulty related to confusion between position and velocity (question 19) was the most resistant to change, and 3% more students had this difficulty in the post-test. In the other question which required interpretation of strobe diagrams of motion (question 20), the normalized gain was 0.25.

**Introductory student shifts in alternate conceptions from pre-test to post-test**

There was only one major shift in alternate conceptions which occurred in more than one question (in questions 5, 11, 18 and 29, which asked students to identify all the distinct forces that act on an object). In the pre-test, many were unaware of contact forces and believed that moving objects are acted upon by a distinct force in the direction of motion, while in the post-test, most students were aware of contact forces. However the alternate conception that there must be a distinct force in the direction of motion was still present.

**Comparison of performance and alternate conceptions of algebra-based with calculus-based students**

In the pre-test, calculus-based students answered every question correctly more frequently than algebra-based students and in the post-test, they answered every question correctly more frequently except for one (question 15). The differences appeared to get larger between these two populations in the post-test compared to the pre-test. In particular, in the pre-test, there were 8 questions in which differences were 20% or larger while in the post-test there were 14 such questions. In addition, the better performance of students in the calculus-based courses compared to the algebra-based courses was context dependent. For example, calculus-based students answered a Newton's third law question better (by 20%) than algebra-based students in the context of question 28 (in which one student was pushing another), but they did not perform better in the other three questions involving the same concept. In fact, in question 15 (a car pushing a truck and accelerating) the percentages of calculus-based and algebra-based students who answered correctly are identical. A similar observation can be made while examining the alternate conception that a constant net force implies constant velocity, in that the better performance of calculus-based students compared to algebra-based students in questions in which this alternate conception can surface is context dependent. In the post-test, algebra-based students consistently answered most questions involving two common alternate conceptions correctly less often (by 20% or more) than calculus-based students. These questions are related to the alternate conceptions that moving objects are acted upon by a distinct force in the direction of motion and that constant net force implies constant velocity. There are other common alternate conceptions which occur on only one question (questions 9, 13, 19 and 20) which are held more strongly by algebra-based students than calculus-based students. We also investigated whether algebra-based students hold different alternate conceptions than calculus-based students. It turns out that was not the case both on the pre-test and on the post-test. On all of the questions, the most common alternate conceptions of algebra-based students and calculus-based students were the same; the difference was that on almost all the questions, more algebra-based students than calculus-based students have these common alternate conceptions both on the pre-test and on the post-test.

Previous studies have found that calculus-based students are more adept than algebra-based students at performing identical tasks that are primarily conceptual [8,57,58,60,61]. The present study corroborates this result because the performance of calculus-based students on the FCI, which is a conceptual assessment, is better than the performance of algebra-based students. It is possible that the better mathematical preparation of calculus-based students helps them develop a better conceptual understanding. In particular, while learning physics, one must process information both about the conceptual and mathematical aspects of physics. A student with a better mathematical preparation can use fewer cognitive resources while engaged in problem solving and learning to process the mathematical aspects, and allocate more cognitive resources to the conceptual aspects. Since working memory is finite, the mathematical facility can reduce the cognitive load [62, 63] and provide more opportunities to build a robust knowledge structure of physics. In contrast, a student lacking the requisite mathematical preparation might spend a significant portion of his/her cognitive resources in processing mathematical information, both while engaged in problem solving and while examining problem solutions. This increased cognitive load can hinder reflection and building of good knowledge structure. Therefore, a better mathematical preparation can help improve conceptual understanding of physics; however, more research is needed to understand the connection between mathematical preparation and conceptual understanding.

**Future research**

In this investigation, instructors and graduate students were asked to identify only one incorrect answer choice for each question on the FCI that they thought would be most commonly selected by introductory students who did not know the correct answer after instruction. However, some of the questions on the FCI contained two common incorrect student answer choices. In addition, instructors were not asked about the difficulty of the questions, which is also an important aspect of the pedagogical content knowledge. Future research will explore these issues and instructors and graduate students will be asked to identify the difficulty level of each question to gauge the extent to which they understand how difficult certain concepts are for introductory students in different contexts. In particular, they will be asked to indicate the percentage of introductory physics students that they think would select each answer choice (incorrect and correct) in a post-test. This future research has the benefit of taking the difficulty of a question into account, and giving the instructors and graduate students the opportunity of indicating more than one incorrect answer choice.

## Acknowledgments

We thank professors F. Reif, J. Levy and R. P. Devaty for helpful discussions and/or feedback on the manuscript.

**Appendix**

**TABLE A1.** Questions on the FCI in which at least 19% of introductory algebra-based students selected one incorrect answer choice in a post-test, percentages of introductory algebra-based physics students who selected each answer choice A through E in a post-test (they were asked to select the correct answer for each question), instructors and graduate students who selected each answer choice A through E (they were asked to select the most common incorrect answer for each question if introductory physics students did not know the correct answer). The first column of the table lists the FCI question numbers and the second column titled > RG shows an "I" when the instructors on average performed better than random guessing, "GS" when the graduate students on average performed better than random guessing; and "I, GS" when both instructors and graduate students performed better than random guessing.

| FCI # | >RG | Intro student choices | | | | | Instructor choices | | | | | Grad student choices | | | | |
|---|---|---|---|---|---|---|---|---|---|---|---|---|---|---|---|---|
| | | A | B | C | D | E | A | B | C | D | E | A | B | C | D | E |
| 2 | I, GS | 44 | 25 | 6 | 21 | 4 | 13 | 53 | 7 | 27 | 0 | 4 | 68 | 0 | 24 | 4 |
| 4 | I, GS | 39 | 1 | 0 | 0 | 60 | 97 | 0 | 0 | 3 | 0 | 84 | 0 | 4 | 12 | 0 |
| 5 | | 3 | 24 | 12 | 44 | 17 | 20 | 0 | 27 | 30 | 23 | 0 | 0 | 32 | 40 | 28 |
| 9 | I, GS | 4 | 20 | 19 | 5 | 53 | 0 | 40 | 54 | 3 | 3 | 12 | 8 | 76 | 0 | 4 |
| 11 | I, GS | 5 | 8 | 48 | 35 | 4 | 17 | 23 | 40 | 0 | 20 | 20 | 20 | 60 | 0 | 0 |
| 12 | I, GS | 1 | 77 | 19 | 2 | 1 | 10 | 3 | 64 | 3 | 20 | 28 | 0 | 68 | 0 | 4 |
| 13 | I | 4 | 11 | 50 | 35 | 0 | 20 | 30 | 47 | 0 | 3 | 36 | 44 | 16 | 0 | 4 |
| 14 | | 19 | 10 | 9 | 61 | 0 | 17 | 57 | 23 | 0 | 3 | 20 | 44 | 36 | 0 | 0 |
| 15 | I | 44 | 7 | 48 | 1 | 0 | 10 | 7 | 60 | 20 | 0 | 12 | 20 | 40 | 28 | 0 |
| 16 | | 73 | 2 | 19 | 2 | 4 | 3 | 17 | 37 | 23 | 20 | 8 | 12 | 16 | 28 | 36 |
| 17 | I, GS | 62 | 24 | 1 | 10 | 3 | 87 | 3 | 0 | 3 | 7 | 72 | 0 | 0 | 16 | 12 |
| 18 | GS | 1 | 28 | 4 | 42 | 25 | 16 | 0 | 47 | 30 | 7 | 4 | 0 | 16 | 52 | 28 |
| 19 | GS | 12 | 3 | 5 | 29 | 51 | 24 | 14 | 21 | 38 | 3 | 8 | 4 | 12 | 76 | 0 |
| 20 | I, GS | 16 | 4 | 27 | 49 | 4 | 7 | 3 | 72 | 3 | 14 | 40 | 4 | 56 | 0 | 0 |
| 21 | I | 7 | 13 | 38 | 9 | 33 | 7 | 40 | 43 | 7 | 3 | 0 | 20 | 44 | 36 | 0 |
| 22 | I | 33 | 45 | 3 | 16 | 2 | 67 | 0 | 7 | 26 | 0 | 28 | 24 | 4 | 40 | 4 |
| 23 | | 15 | 39 | 18 | 23 | 5 | 20 | 7 | 47 | 23 | 3 | 20 | 4 | 28 | 24 | 24 |
| 24 | I | 70 | 2 | 22 | 2 | 5 | 7 | 3 | 70 | 0 | 20 | 0 | 16 | 68 | 12 | 4 |
| 25 | I | 3 | 9 | 23 | 53 | 12 | 10 | 17 | 0 | 57 | 16 | 24 | 24 | 0 | 44 | 8 |
| 26 | I | 41 | 32 | 3 | 9 | 14 | 60 | 27 | 0 | 13 | 0 | 52 | 32 | 0 | 12 | 4 |
| 27 | I, GS | 26 | 13 | 58 | 2 | 0 | 63 | 27 | 0 | 7 | 3 | 68 | 16 | 4 | 4 | 8 |
| 28 | GS | 1 | 2 | 6 | 32 | 59 | 7 | 45 | 10 | 38 | 0 | 4 | 36 | 8 | 52 | 0 |
| 29 | | 2 | 71 | 3 | 23 | 1 | 24 | 3 | 0 | 28 | 45 | 20 | 8 | 0 | 28 | 44 |
| 30 | I, GS | 3 | 10 | 26 | 2 | 59 | 3 | 0 | 0 | 3 | 94 | 4 | 4 | 0 | 4 | 88 |

| | |
|---|---|
| x | Correct answer |
| x | x > 33 – strong alt conception (more than 1/3 of intro students chose it) |
| x | 19 ≤ x ≤ 33 – medium level alternate conception |

**TABLE A2.** Questions on the FCI in which at least 19% of introductory algebra-based students selected one incorrect answer choice in a post-test, percentages of introductory algebra-based students who answer each question correctly (% intro. alg. correct), normalized gain (Intro. alg. norm. gain), maximum possible FCI related PCK score (max. pos. PCK score), Average FCI related PCK score of instructors (Ins. avg. PCK score) and graduate students (GS avg. PCK score), percentage of maximum possible score of the instructors' average FCI related PCK score (Ins. % max score) and of the graduate students average FCI related PCK score (GS % max score).

| FCI item # | % intro. alg. correct | Intro. alg. norm. gain | Min. pos. PCK score | Max. pos. PCK score | Instructors | | Graduate students | |
|---|---|---|---|---|---|---|---|---|
| | | | | | Ins. avg. PCK score | Norm Ins. avg. PCK score | GS avg. PCK score | Norm GS avg. PCK score |
| 2 | 44 | 0.23 | 0.04 | 0.25 | 0.19 | 71 | 0.22 | 86 |
| 4 | 60 | 0.46 | 0 | 0.39 | 0.38 | 97 | 0.33 | 85 |
| 5 | 24 | 0.16 | 0.03 | 0.44 | 0.21 | 44 | 0.26 | 56 |
| 9 | 53 | 0.17 | 0.04 | 0.20 | 0.18 | 88 | 0.17 | 81 |
| 11 | 35 | 0.24 | 0.04 | 0.48 | 0.23 | 43 | 0.31 | 61 |
| 12 | 77 | 0.43 | 0.01 | 0.19 | 0.12 | 61 | 0.13 | 67 |
| 13 | 35 | 0.27 | 0 | 0.50 | 0.27 | 54 | 0.14 | 28 |
| 14 | 61 | 0.39 | 0 | 0.19 | 0.11 | 58 | 0.11 | 58 |
| 15 | 44 | 0.25 | 0 | 0.48 | 0.30 | 63 | 0.21 | 44 |
| 16 | 73 | 0.40 | 0.02 | 0.19 | 0.09 | 41 | 0.05 | 18 |
| 17 | 24 | 0.17 | 0.01 | 0.62 | 0.54 | 87 | 0.47 | 75 |
| 18 | 28 | 0.19 | 0.01 | 0.42 | 0.16 | 37 | 0.30 | 71 |
| 19 | 51 | -0.06 | 0.03 | 0.29 | 0.15 | 46 | 0.24 | 81 |
| 20 | 49 | 0.25 | 0.04 | 0.27 | 0.21 | 74 | 0.22 | 78 |
| 21 | 33 | -0.02 | 0.07 | 0.38 | 0.23 | 52 | 0.23 | 52 |
| 22 | 45 | 0.22 | 0.02 | 0.33 | 0.26 | 77 | 0.16 | 45 |
| 23 | 39 | 0.14 | 0.05 | 0.23 | 0.17 | 67 | 0.15 | 56 |
| 24 | 70 | 0.20 | 0.02 | 0.22 | 0.16 | 70 | 0.16 | 70 |
| 25 | 23 | 0.13 | 0.03 | 0.53 | 0.34 | 62 | 0.27 | 48 |
| 26 | 14 | 0.11 | 0.03 | 0.41 | 0.34 | 82 | 0.33 | 79 |
| 27 | 58 | 0.09 | 0 | 0.26 | 0.20 | 77 | 0.20 | 77 |
| 28 | 59 | 0.6 | 0.01 | 0.32 | 0.14 | 42 | 0.18 | 55 |
| 29 | 71 | 0.50 | 0.02 | 0.23 | 0.07 | 24 | 0.07 | 24 |
| 30 | 26 | 0.16 | 0.02 | 0.59 | 0.55 | 93 | 0.53 | 89 |

| # | Question in which there was a medium level alternate conception |
|---|---|
| # | Question in which there was a strong student alternate conception |
| x | Ins./grad students' FCI related PCK score is less than 50% of maximum possible |
| x | Ins./GS score FCI rel. PCK score is between 50% and 67% of maximum possible |
| x | Ins./GS FCI related PCK score is more than 67% of maximum possible |

**TABLE A3.** Percentages of algebra-based introductory physics students who selected each answer choice for each item on the FCI when it was given in a pre-test and in a post-test and normalized gain (Norm. gain) on each item on the FCI. The percentages on the pre-test are based on data from 601 students taught by two different instructors in two different semesters and the percentages on the post-test are based on data from 899 students taught by 4 different instructors over several years. The green shaded boxes indicate correct answers. All the courses were taught in a traditional manner which did not incorporate PER based teaching strategies.

| FCI item # | Pre-test algebra | | | | | Post-test algebra | | | | | Norm. gain |
|---|---|---|---|---|---|---|---|---|---|---|---|
| | A | B | C | D | E | A | B | C | D | E | |
| 1 | 13 | 6 | 53 | 25 | 4 | 10 | 4 | 78 | 8 | 1 | 0.53 |
| 2 | 27 | 21 | 7 | 37 | 8 | 44 | 25 | 6 | 21 | 4 | 0.23 |
| 3 | 31 | 16 | 40 | 3 | 10 | 15 | 13 | 63 | 5 | 5 | 0.38 |
| 4 | 73 | 0 | 0 | 1 | 26 | 39 | 1 | 0 | 0 | 60 | 0.46 |
| 5 | 4 | 10 | 31 | 25 | 29 | 3 | 24 | 12 | 44 | 17 | 0.16 |
| 6 | 25 | 68 | 5 | 2 | 0 | 16 | 79 | 3 | 1 | 1 | 0.33 |
| 7 | 17 | 57 | 9 | 5 | 11 | 12 | 74 | 6 | 3 | 4 | 0.40 |
| 8 | 20 | 47 | 1 | 13 | 18 | 14 | 66 | 0 | 8 | 11 | 0.36 |
| 9 | 4 | 26 | 20 | 6 | 43 | 4 | 20 | 19 | 5 | 53 | 0.17 |
| 10 | 54 | 1 | 11 | 19 | 15 | 71 | 2 | 7 | 13 | 7 | 0.36 |
| 11 | 7 | 31 | 45 | 14 | 3 | 5 | 8 | 48 | 35 | 4 | 0.24 |
| 12 | 1 | 59 | 32 | 5 | 2 | 1 | 77 | 19 | 2 | 1 | 0.43 |
| 13 | 4 | 21 | 64 | 12 | 0 | 4 | 11 | 50 | 35 | 0 | 0.27 |
| 14 | 35 | 18 | 11 | 36 | 0 | 19 | 10 | 9 | 61 | 0 | 0.39 |
| 15 | 25 | 10 | 61 | 3 | 0 | 44 | 7 | 48 | 1 | 0 | 0.25 |
| 16 | 55 | 3 | 37 | 4 | 1 | 73 | 2 | 19 | 2 | 4 | 0.40 |
| 17 | 60 | 8 | 1 | 22 | 9 | 62 | 24 | 1 | 10 | 3 | 0.17 |
| 18 | 2 | 12 | 14 | 27 | 46 | 1 | 28 | 4 | 42 | 25 | 0.19 |
| 19 | 14 | 3 | 3 | 26 | 54 | 12 | 3 | 5 | 29 | 51 | -0.06 |
| 20 | 19 | 6 | 36 | 32 | 8 | 16 | 4 | 27 | 49 | 4 | 0.25 |
| 21 | 7 | 21 | 23 | 14 | 35 | 7 | 13 | 38 | 9 | 33 | -0.02 |
| 22 | 37 | 30 | 4 | 26 | 2 | 33 | 45 | 3 | 16 | 2 | 0.22 |
| 23 | 16 | 29 | 21 | 28 | 7 | 15 | 39 | 18 | 23 | 5 | 0.14 |
| 24 | 63 | 1 | 25 | 3 | 6 | 70 | 2 | 22 | 2 | 5 | 0.20 |
| 25 | 3 | 8 | 12 | 58 | 19 | 3 | 9 | 23 | 53 | 12 | 0.13 |
| 26 | 41 | 42 | 4 | 11 | 3 | 41 | 32 | 3 | 9 | 14 | 0.11 |
| 27 | 31 | 13 | 54 | 1 | 0 | 26 | 13 | 58 | 2 | 0 | 0.09 |
| 28 | 0 | 6 | 8 | 61 | 24 | 1 | 2 | 6 | 32 | 59 | 0.46 |
| 29 | 15 | 42 | 1 | 37 | 4 | 2 | 71 | 3 | 23 | 1 | 0.50 |
| 30 | 1 | 7 | 12 | 1 | 79 | 3 | 10 | 26 | 2 | 59 | 0.16 |
| | | | | | | | Avg. normalized gain | | | | 0.26 |

| | |
|---|---|
| x | Correct answer |
| x | x > 33 – strong alternate conception (more than 1/3 of intro students chose it) |
| x | 19 ≤ x ≤ 33 – medium level alternate conception. |

**TABLE A4.** Percentages of calculus-based introductory physics students who selected each answer choice for each item on the FCI when it was given in a pre-test and in a post-test. The percentages on the pre-test are based on data from 364 students taught by three different instructors over several semesters and the percentages on the post-test are based on data from 296 students taught by two different instructors during two different semesters. The green shaded boxes indicate correct answers. All the courses were taught in a traditional manner which did not incorporate PER based teaching strategies.

| FCI item # | Pre-test calculus | | | | | Post-test calculus | | | | | Norm. gain |
|---|---|---|---|---|---|---|---|---|---|---|---|
| | A | B | C | D | E | A | B | C | D | E | |
| 1 | 8 | 3 | 82 | 6 | 1 | 7 | 1 | 86 | 6 | 1 | 0.24 |
| 2 | 39 | 24 | 4 | 26 | 7 | 55 | 27 | 4 | 15 | 1 | 0.25 |
| 3 | 17 | 8 | 66 | 4 | 5 | 15 | 4 | 76 | 3 | 2 | 0.30 |
| 4 | 59 | 1 | 0 | 0 | 39 | 23 | 0 | 1 | 0 | 76 | 0.60 |
| 5 | 5 | 23 | 20 | 27 | 25 | 3 | 47 | 13 | 29 | 9 | 0.31 |
| 6 | 17 | 79 | 3 | 1 | 1 | 12 | 86 | 2 | 1 | 0 | 0.32 |
| 7 | 9 | 76 | 4 | 3 | 7 | 7 | 85 | 3 | 1 | 5 | 0.39 |
| 8 | 11 | 62 | 0 | 12 | 17 | 8 | 74 | 0 | 8 | 10 | 0.33 |
| 9 | 2 | 20 | 20 | 6 | 53 | 1 | 9 | 14 | 3 | 73 | 0.44 |
| 10 | 70 | 3 | 6 | 12 | 8 | 91 | 1 | 3 | 3 | 3 | 0.69 |
| 11 | 8 | 16 | 42 | 30 | 4 | 6 | 2 | 28 | 61 | 3 | 0.44 |
| 12 | 0 | 84 | 14 | 2 | 0 | 0 | 90 | 10 | 0 | 0 | 0.41 |
| 13 | 4 | 12 | 51 | 33 | 0 | 3 | 5 | 31 | 61 | 0 | 0.42 |
| 14 | 21 | 12 | 5 | 63 | 0 | 14 | 9 | 5 | 72 | 0 | 0.25 |
| 15 | 29 | 6 | 63 | 2 | 0 | 42 | 5 | 52 | 0 | 1 | 0.19 |
| 16 | 72 | 1 | 23 | 2 | 3 | 86 | 2 | 7 | 1 | 4 | 0.50 |
| 17 | 63 | 21 | 1 | 12 | 4 | 46 | 44 | 2 | 6 | 2 | 0.29 |
| 18 | 3 | 27 | 9 | 34 | 29 | 0 | 55 | 2 | 33 | 11 | 0.38 |
| 19 | 11 | 3 | 3 | 20 | 62 | 8 | 2 | 3 | 12 | 75 | 0.35 |
| 20 | 11 | 4 | 22 | 54 | 6 | 9 | 2 | 16 | 71 | 3 | 0.36 |
| 21 | 6 | 11 | 26 | 11 | 46 | 4 | 6 | 23 | 11 | 57 | 0.20 |
| 22 | 27 | 51 | 2 | 19 | 1 | 20 | 67 | 2 | 11 | 1 | 0.34 |
| 23 | 9 | 48 | 16 | 24 | 3 | 5 | 64 | 13 | 14 | 3 | 0.32 |
| 24 | 74 | 1 | 18 | 2 | 4 | 87 | 1 | 9 | 1 | 3 | 0.48 |
| 25 | 2 | 9 | 29 | 51 | 10 | 3 | 6 | 51 | 38 | 2 | 0.31 |
| 26 | 35 | 29 | 1 | 14 | 20 | 20 | 28 | 3 | 7 | 42 | 0.28 |
| 27 | 22 | 7 | 66 | 4 | 1 | 14 | 6 | 75 | 3 | 1 | 0.28 |
| 28 | 2 | 4 | 6 | 45 | 44 | 1 | 3 | 3 | 20 | 72 | 0.51 |
| 29 | 6 | 61 | 2 | 27 | 3 | 1 | 83 | 1 | 13 | 2 | 0.56 |
| 30 | 3 | 7 | 30 | 1 | 59 | 3 | 7 | 44 | 1 | 45 | 0.20 |
| | | | | | | | | Avg. normalized gain | | | 0.36 |

| | |
|---|---|
| x | Correct answer |
| x | x > 33 – strong alternate conception (more than 1/3 of intro students chose it) |
| x | 19 ≤ x ≤ 33 – medium level alternate conception |